%% file: ConspConsRev.tex
\documentclass[12pt]{article}
\usepackage{comment}

\input{preambleR}

\title{The Joneses Visit an Economics Lab}

\author{
  Mikhail Freer\thanks{
  University of Barcelona.  Email: mikhail.freer@ub.edu} 
\and
  Daniel Friedman\thanks{
  University of Essex and University of California Santa Cruz. Email: dan@ucsc.edu} 
\and
  Christian Ghiglino \thanks{
  University of Essex. Email: cghig@essex.ac.uk}
\and 
  Elke Weidenholzer \thanks{
  University of Essex. Email: elke.weidenholzer@essex.ac.uk}
  \thanks{
We are grateful to Rojo Arjona, Yann Bramoullé, Antonio Cabrales, Ahrash Dianat, 
Charles Figuières, Sanjeev Goyal, Natalia Lazzati, Gerelt Tserenjigmid, and Kristian Lopez Vargas
for helpful comments,  
and to audiences at 
Network Science in Economics April 2026 Miami, 
UCSC Brown Bag workshop April 2026,
Bristol Networks Workshop May 2026,
25th LAGV Conference Aix-en-Provence June 2026, and
Birmingham Economic Theory Workshop June 2026.
We gratefully acknowledge funding from the Philipp Leverhulme Trust (Leverhulme Research Leadership Award Competition and Competitiveness) and Ethics approval under ETH2122-1491.}
}

\date{16 September 2026}

\begin{document}

\maketitle

\begin{singlespace}

\begin{abstract}
Existing literature offers persuasive evidence that individuals care about how their consumption compares to that of peers,
and proposes a large variety of explanatory models. 
The present paper proposes a common framework for many of those models,
and compares their ability to predict behavior in a laboratory experiment. 
We find evidence of \emph{Keeping up with the Joneses} motivations but also find that conspicuous consumption is enhanced by  \emph{Veblen} motivations arising from peers' ability to observe one's own choice.
Among the seven quasi-linear preference models we compare, our data are best explained by a model
that contrasts envy and pride (upward vs downward comparisons) using a value function borrowed from Prospect Theory.


\end{abstract}


\noindent \textbf{Preliminary Version.} Please do not circulate without permission.


\end{singlespace}


\newpage
\setcounter{page}{1}

\maketitle

\section{Introduction}\label{NewIntro}

Thorstein \cite{veblen1899thetheory} famously argued that people consume conspicuously to display social status or identity. 
Using recent jargon, his idea is that social image concerns may shape revealed preferences for highly visible goods that 
signal unobservable traits to an audience of peers.

Self-image concerns may also be important. 
According to the famous \emph{Keeping up with the Joneses} effect (\cite{duesenberry1949income}, \cite{abel1990asset}),  
people compare themselves to others in terms of proxies for income, ability, or success. 
Even in the absence of an audience, such comparisons can generate emotions such as envy or pride.

For both sorts of concerns, 
what matters is not the absolute level of consumption, but rather one's relative position within the observed distribution of peers' consumption choices. 
Existing literature offers an impressive variety of theoretical models for how relative position matters. 
Some emphasize 
comparison to the mean, while others posit ordinal rank-based preferences and/or envy–pride comparisons.

Empirical discrimination among such models is difficult. Field data often suffers from limited observability, institutional confounds, and endogenous peer-group formation. 
Relevant experiments are rare, perhaps because it is not obvious how to elicit preferences for conspicuous consumption in the lab.

The present paper aims to shed light on such matters by (a) proposing a common framework for many theoretical models of conspicuous consumption, and (b) comparing the power of those models to predict behavior observed in a new laboratory experiment. 

Following a literature survey later in this Introduction, we proceed as follows.
Section \ref{sec:theory} embeds several prominent models of conspicuous consumption within a unified quasi-linear framework. Players allocate endowed income between ordinary private consumption (money, i.e., general purchasing power)
and a visible good whose payoff derives solely from social comparison. 
Within that framework, we compare various models of social comparison including a simple mean-based model, a conformity model, and an ordinal rank-dependent model.
We also consider cardinal models in which upward (``envy'') and downward (``pride'') comparisons may affect utility asymmetrically. 
Finally, inspired by Prospect Theory and related empirical evidence, 
we explore whether an S-shaped value function can improve explanatory power. 
We derive contrasting predictions across those models in terms of
best-responses and comparative statics.

Section \ref{sec:design} describes a laboratory experiment designed to test those predictions.
Participants earn income through a real-effort task and then choose how much of it to spend on a purely positional good --- digital ``decals'' that are seen by peers but yield no material payoff. 
The experiment proceeds in repeated rounds with fixed peer groups.
In later rounds, participants have the unilateral opportunity to revise their own decal purchase
after seeing peers' purchases, i.e., to best-respond as last mover. 
Treatments vary the degree of visibility in order to isolate self-image (\emph{``Keeping up with the Joneses''}) concerns from 
social-image concerns (\emph{``Veblen''}). 

Section \ref{sec:results} documents several robust patterns in the laboratory data. 
First, 
conspicuous consumption (i.e., decal expenditure, referred to henceforth as CC) increases roughly one-for-one with peers' mean CC, indicating a high sensitivity to social comparison. This relationship holds most clearly in the  treatment where others’ behavior is directly observed and players can best respond. 
Players' CC is largely independent of their earnings, consistent with our quasi-linear framework and ruling out models with strong income effects.

Second, visibility significantly affects behavior. In our Veblen treatment, where players observe each other's identity, CC is higher and more persistent than in the Jones treatment, where peers are anonymous. 
CC is considerably larger in later rounds, when participants can revise their decisions. 
That CC increase is mainly at the extensive margin: 
a larger fraction of participants choose positive levels of CC in those later rounds.

Third, our data  are inconsistent with several standard models. Somewhat surprisingly, we reject rank-based models. 
Those models predict that individuals target a specific rank and so should choose CC just above a peer’s choice, 
but such behavior is rare in our data. 
The Conformity model predicts that CC choices will become smaller and less dispersed, but no such time trends appear in our data. 
The simple average model offers no meaningful predictions in our setting. 
Linear envy--pride models successfully capture the response to peers' mean CC but fail  to capture responses to dispersion of peers' CC. 

The model that appears best able to track our data uses pairwise comparisons and the S-shaped value function adapted from Prospect Theory. 
This model correctly predicts (i) a roughly one-for-one response of CC to shifts in peers’ mean consumption, (ii) a tendency towards diminishing sensitivity to peers' consumption further from one's own, and (iii) interior choices that vary systematically with peers' chosen levels of CC rather than (as some other models predict) a predominance of corner solutions.

Section \ref{sec:conclude} concludes, and Appendices report supplementary data analysis, mathematical details, and Instructions to Participants. 

\bigskip

\noindent
\textbf{Existing Literature.}

Our paper relates to a vast literature on social preferences, which posits that individuals care about others’ consumption. 
Various strands of that literature formalize concepts such as altruism, fairness, reciprocity, and inequality aversion. 
\cite{FehrCharness2025} review a prominent strand following \cite{FehrSchmidt1999},
much of it based on laboratory experiments.
That strand highlights pairwise payoff differences, and often emphasizes an asymmetry between upward and downward differences.

Another strand of literature studies \emph{Keeping up with the Joneses} concerns, where others’ choices affect utility by shaping reference points, habits, or aspirations. 
Empirical work typically emphasizes mean-based comparisons in total consumption or income, e.g., \cite{de2020consumption}. 
\cite{drechsel2014consumption} and \cite{bertrand2016trickle} also focus on 
comparisons to the mean but assume that individuals only look upwards, a strong version of asymmetry. 
Extensive empirical evidence suggests that upward income comparisons weigh more heavily than downward comparisons (\cite{ferrer2005income, vendrik2007happiness, senik2009direct} and \cite{leites2022effect}). 

In a different strand following \cite{veblen1899thetheory}, the value of consumption derives from its social interpretation and observability becomes central. 
\cite{charles2009conspicuous} show that Black and Hispanic households allocate more spending to visible goods than otherwise similar White households, consistent with status signaling in environments where income is imperfectly observed. \cite{BrownBulteZhang2011} document status competition in Chinese villages, where publicly observable consumption proxies generate rank-based social comparisons within small reference groups. 

Field-experimental evidence in \cite{bursztyn2018status} demonstrates that demand for status goods (platinum credit cards) arises even when functional benefits are held constant, and that usage is concentrated in socially visible contexts. Related experimental work shows that visibility alone can generate costly signaling even under anonymity. \cite{ClingingsmithSheremeta2018} study rank-based status competition following performance feedback, while \cite{BanuriNguyen2023} show that observability increases both conspicuous consumption and borrowing when individuals seek to signal status despite budget constraints.

A distinct literature studies social-image concerns rather than consumption per se. In \cite{Bernheim1994}, conformity arises because individuals derive utility from social approval, and subsequent experimental work such as \cite{AndreoniBernheim2009}. \cite{petrishcheva2023loss} show that individuals exhibit loss aversion in social image, engaging in dishonest behavior to avoid public rank losses. 

Reference points play a central role in Prospect Theory (PT), following \cite{kahneman1979prospect} and \cite{tversky1992advances}. In the dimension relevant for our analysis, PT is characterized by three closely related features: outcomes are evaluated relative to a reference point, losses relative to that reference point are weighted more heavily than comparable gains, and sensitivity to gains and losses decreases with their distance from the reference point. These properties give rise to the familiar S-shaped value function and, in particular, to the concept of loss aversion (\cite{kahneman1979prospect}, \cite{tversky1992advances}, \cite{loewenstein1989social}, \cite{petrishcheva2023loss}, \cite{brown-metaanalysis}). 
Although PT was initially developed to study decision-making under risk, its applications have since expanded well beyond that domain; see \cite{thaler1980toward} and \cite{barberis2013thirty} for surveys. To our knowledge, however, these three PT features have not previously been incorporated jointly into a model of conspicuous consumption, much less tested in that context. 


Conspicuous consumption has been shown to arise in large population status games where individuals care explicitly about their rank (\cite{frank1985demand}, \cite{hopkins2004running}). \cite{hopkins2023cardinal} extends 
the ordinal model of \cite{hopkins2004running} so that players care, potentially asymmetrically, about their cardinal positions in the status distribution. Asymmetric comparisons in a rank model are introduced in \cite{frank2014expenditure}, where individuals compare to reference levels determined by nearby higher-ranked players, rather than the average consumption of all higher-ranked individuals. 

On the other hand, in the  \emph{Keeping up with the Joneses} literature, individuals typically do compare their consumption or income to economy wide references given by averages (\cite{duesenberry1949income}, \cite{abel1990asset}, \cite{campbell1999force}, \cite{ljungqvist2000tax}). \cite{friedman2008conspicuous} introduce asymmetries in a model of diffuse social effects. They consider a continuum of consumers with identical preferences, where upward (envy) comparisons of own conspicuous consumption to individual peers' CC are weighted differently than downward comparisons (pride).

\cite{zenoupeer} develop a general theory of peer effects in which the relevant social reference can depend on different parts of the peer distribution. 
In a recent strand of theoretical literature, individuals compare their own consumption or action with that of a small set of neighbors,
connected to the literature on games played on networks; see \cite{bramoulle2016games} and \cite{jackson2015games} for reviews.
See \cite{ushchev2020social} for a network game with conformism, where players compare their own effort to the average effort among their network peers, 
and for status games see \cite{ghiglino2010keeping}, \cite{immorlica2017social} and \cite{langtry2022keeping}. 

In \cite{immorlica2017social}, a player's utility depends linearly on a weighted sum of differences between own costly action and the actions of neighbors taking a higher action, an assumption akin to an extreme form of loss aversion. There is no social reference point and individual incomes play no role.  \cite{langtry2022keeping} modifies \cite{immorlica2017social}'s framework and assumes that players form a social reference point based on their neighbors' consumption. \cite{sadler2023games} extend \cite{immorlica2017social} in that players compete for status and simultaneously choose their connections which provide a fixed benefit. \cite{lopez-pintado2021far} study a game of effort provision, rather than status: 
players gain an extra utility when producing an outcome above a \textquotedblleft comparison  threshold\textquotedblright\ derived from the outcomes of their reference group. Finally, in \cite{bramoulle2024status} consumers allocate heterogeneous incomes across two substitutable goods, but their preferences exhibit loss aversion in regard to a reference point given by neighbors averages.



\section{Testable Theory}\label{sec:theory}
In this section, we build a framework that enables comparisons of various proposed models of conspicuous consumption preferences.
The models assume that players can observe the distribution of peers' consumption of a relevant good,
but make no assumption about how visible the player's own consumption is to peers.
That is, the models might represent either \emph{Joneses} preferences or \emph{Veblen} preferences, possibly with different fitted parameters.
We spotlight how the models differ in terms of potentially observable implications. 

\subsection{A common framework}
Consider a fixed set of players $i=0,1, ..., N$, denoting a focal player by $i=0.$ 
Let $y_i \geq 0$ denote player $i$'s expenditures on conspicuous consumption, and let
$m_i$ denote her remaining purchasing power. The vector $y_{-i}$ denotes the other players' conspicuous consumption, e.g., $y_{-0} = (y_1, ...,y_n)$ is conspicuous consumption by the focal player's peers.
Without loss of generality, we re-index the peers so that $y_1 \geq y_2 \geq ... \geq y_N\geq 0$, with the notational convention that $y_{N+1} = 0$.

The focal player's \emph{rank} $r$ is $N+1$
if $y_0=0$ and otherwise
is the unique positive integer such that $y_0 \in (y_{r}, y_{r-1}]$. 
Appendix B explains how this convention discourages the focal player from seeking ties.
\emph{Mean conspicuous consumption} by the focal player's peers is
$\Bar{y}=\frac{1}{N}\sum_{i=1}^N y_i$. 
If the function $f$ has an isolated discontinuity at $x,$ then the right- (resp. left-) hand limit is denoted by 
$f(x+)=lim_{z \searrow x}f(z)$ 
(resp. $f(x-)=lim_{z \nearrow x}f(z)$).

Most of the models we consider assume quasi-linear preferences of the general form 
\begin{equation}\label{eq:qlinpref} 
U(m_0,y_0, y_{-0}) = m_0 +\phi(y_0, y_{-0}).
\end{equation}
The subutility function  $\phi$ for conspicuous consumption is (unless otherwise stated) piecewise differentiable, increasing in own conspicuous consumption $y_0$, and decreasing in each component of peers' consumption $y_{-0}= (y_1,..., y_N)$. 

\subsection{Candidate Models}
We shall investigate the following models of conspicuous consumption.\\

\noindent
\textbf{Simple average model (SAM)}: $\phi(y_0, y_{-0}) = y_0 - c\Bar{y} $,
with parameter $c>0$.
Such models appear in some of the macroeconomics literature, e.g., 
\cite{carroll1997comparison}, \cite{ljungqvist2000tax}, \cite{clark1996satisfaction}; see also \cite{abel1990asset}.\\

\noindent
\textbf{Conformity}: $\phi(y_0, y_{-0}) = -c(y_0 - \Bar{y})^2$,
where $c=c_p>0$ when $y_0 \geq \Bar{y}$ and $c=c_e \geq c_p$ when $y_0 < \Bar{y}$.
This version allows asymmetry between upwards and downwards comparisons to mean CC, but deviations from the mean are never desirable.
Consequently our monotonicity assumptions 
fail in this model in some regions. Relevant literature includes \cite{Bernheim1994}, \cite{akerlof1997social}, and \cite{bisin2011economics}.\\

\noindent
\textbf{Rank-dependent}: $\phi(y_0, y_{-0}) = C(r)$,
where $r$ is the focal player's rank and $C: \mathcal{N} \rightarrow \Re$ is a decreasing function defined on positive integers. That is, subutility $\phi$ does not depend on cardinal values of conspicuous consumption but only on where the focal player ranks in the distribution, and rank 1 is best, followed by rank 2, etc. See for example
\cite{frank1985choosing}, \cite{hopkins2004running}, \cite{robson1992status}, and \cite{moldovanu2007contests}.\\

\noindent
\textbf{Peerwise Envy-Pride (PEP)}: $\phi(y_0, y_{-0}) = \frac{c_e}{N} \sum_{i=1}^{r-1} [y_0 -y_i ] + \frac{c_p}{N} \sum_{i=r}^N [y_0 -y_i]$.
Here envy is disutility from shortfalls ($[y_0 -y_i]<0$) in upward comparisons to each peer with greater conspicuous consumption, while pride boosts utility from downward comparisons ($[y_0 -y_i ]>0)$.
Related literature typically assumes $c_e \geq c_p \geq 0$. 
The $1/N$ normalization reflects the presumption that peer group size shouldn't matter much.
This specification is related to inequality-aversion models following \cite{FehrSchmidt1999}, and to the relative-deprivation literature following \cite{stark1984rural}.\\

\noindent
\textbf{Mean Envy-Pride (MEP)}: $\phi(y_0,{y}_{-0})= c[y_0-\bar{y}]$,
where $c=c_p>0$ when $y_0 \geq \Bar{y}$ and $c=c_e \geq c_p$ when $y_0 < \Bar{y}$.
Here envy and pride arise from comparison to the mean of peers' conspicuous consumption, rather than to individual peers' consumption levels. See, e.g.,  \cite{bolton2000erc} or \cite{bramoulle2024status}. \\

One might question the linearity assumptions in MEP and PEP for large upward and downward comparisons. For example, when above $\bar{y}$ in MEP, the marginal benefit of increasing $y_0$ is always $c_p$, but the marginal benefit might be far greater when just slightly above $\bar{y}$ than when far above it.
To incorporate such diminishing sensitivity,
our two final models use a standard device, the S-shaped value function from Prospect Theory. 

\begin{figure}[!htb]
    \centering
    \begin{tikzpicture}
        \begin{axis}[
            axis lines = middle,
            xlabel = {CC gap $x=(y_0 - y_i)$},
            ylabel = {Value $V(x)$},
            ticks = none,
            xmin = -4, xmax = 4,
            ymin = -2, ymax = 2,
            every axis x label/.style={at={(ticklabel* cs:1)}, anchor=north},
            every axis y label/.style={at={(ticklabel* cs:1)}, anchor=east},
        ]
            \addplot[domain=0:4, blue, ultra thick, samples=100] 
            {1-exp(-0.6*x)}; 
            \node[blue] at (axis cs: 1.5, 1.1) {Gains = downward comps};

            \addplot[domain=-4:0, red, ultra thick, samples=100] 
            {-1+exp(1.8*x)}; 
            \node[red] at (axis cs: -1.5, -1.2) {Losses = upward comps};

            
        
        \end{axis}
\end{tikzpicture}
\caption{Loss aversion in PT,
for piecewise CARA value function $V$ in equation (\ref{eq:PTvalfn}). The slope of the tangent to the blue (red) curve at $x=0$ is $c_p$ (resp. $c_e$). In the example shown,  $(c_p, c_e)=(0.6, 1.8)$. }    
   \label{fig:PTvalfn}
\end{figure}
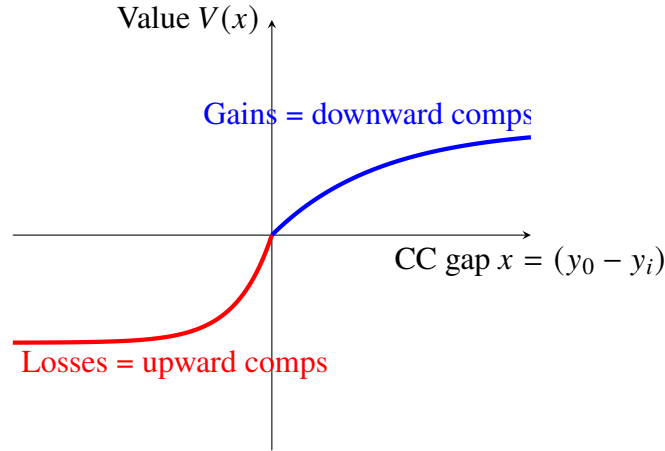

As illustrated in Figure \ref{fig:PTvalfn},
the value function is $V(x)$ is defined over positive (``gains'') and negative (``losses'') deviations $x$ of focal conspicuous consumption $y_0$ from a specified reference point. 
That reference point is 
each peer's $y_i$ in
the PT-PEP model defined below, while the PT-MEP model uses peers' mean $\bar{y}$.

The function $V$ is continuous, strictly increasing and
satisfies $V(0)=0$. 
It is twice differentiable everywhere except 
at $x=0$, where it has a kink, viz.  $V'(0-)>V'(0+)$, representing loss aversion or, in our setting, 
the gap between envy and pride. 
It is S-shaped in that $xV''(x)<0$ for $x\neq0.$
The standard parametric example is CARA:
\begin{equation}\label{eq:PTvalfn}
    V(x) =
    \begin{cases}
      1-e^{-c_p x} & \mbox{ if } \ x \geq 0 \\
      -1+ e^{c_e x} & \mbox{ if } \ x<0.
    \end{cases}
  \end{equation}
When $0<c_p<c_e$ we have the desired kink at zero, 
since $V'(0+) = c_p$ and 
$V'(0-) = c_e$, as in Figure \ref{fig:PTvalfn}.
We have the desired S-shape since
the marginal benefit  $V'(x)= c_p e^{-c_p x}>0 $ is decreasing for $x>0$ and $V'(x)=c_e e^{c_e x}>0$ is increasing for $x<0$.  

We now specify our last two models.

\noindent
\textbf{Prospect Theory - Mean Envy-Pride (PT-MEP)}: $\phi(y_0,{y}_{-0})= V(y_0-\bar{y})$.\\
Note that the piecewise linear model MEP introduced earlier is a first order approximation of (\ref{eq:PTvalfn}) for small upward and downward deviations
$x=y_0 - \Bar{y}$. 

\bigskip
\noindent
\textbf{Prospect Theory - Peerwise Envy-Pride (PT-PEP)}: $\phi(y_0,{y}_{-0})=
 \frac{1}{N}\sum_{i=1}^N V(y_0 - y_i)$.
Likewise, the PEP model is a first order approximation for small deviations
$x=y_0 - y_i$. 

\subsection{Analysis of Models}\label{ssec:ModAnal}

Many of our models' observable  implications arise from their best response functions, that is, their solutions to 
\begin{equation}\label{eq:brprob}
\max_{y_0 \in [0, w_0]} U(m_0, y_0, y_{-0}) \ \ s.t. \ \  m_0 = w_0 - y_0.
\end{equation}
Thus, given her available income $w_0$, the focal agent chooses CC expenditure $y_0$ that best responds to peers' CC choices $ y_{-0}$, leaving remaining purchasing power $m_0 = w_0 - y_0$.
For the quasi-linear specification in equation  (\ref{eq:qlinpref}) above, 
we have $U=w_0 - y_0 + \phi(y_0, y_{-0})$.
Interior solutions satisfy the first order condition (FOC) $0=\frac{dU}{dy_0}=\frac{\partial U}{\partial m_0}\frac{d m_0}{dy_0} + \frac{\partial U}{\partial y_0}\frac{d y_0}{dy_0} + 0 = -1 + \phi_1 + 0,$ which we rewrite as
\begin{equation}\label{eq:foc4ql}
1= \phi_1 \equiv \frac{\partial}{\partial y_0}\phi(y_0, y_{-0}). 
\end{equation}
Here 1 is the marginal cost (or price) of CC while $\phi_1$
is the marginal benefit. 
Intuitively, $\phi_1 -1$ is the net incentive to increase own CC (or to decrease it when $\phi_1 -1<0$).

Note that income effects are zero at interior solutions, since $w_0$ does not appear in equation (\ref{eq:foc4ql}), although they resurface at corner solutions $y_0=w_0$. 
By contrast, as shown in Appendix \ref{ssec:AppMultM},
interior solutions have substantial income effects in specifications   
where 
$\phi$ and $m$ interact multiplicatively.
Examples in existing literature include \cite{ghiglino2010keeping}, where $\phi$ is essentially as in the MEP model, and \cite{hopkins2004running}, where $\phi$ is rank dependent.

Our first testable prediction is that the quasi-linear framework is adequate: 
\begin{prediction} (QLin).
Consistent with equation (\ref{eq:qlinpref}), income effects estimated from interior best responses will be near zero. 
\end{prediction}

Assuming that Prediction A is not rejected, we now look for ways to distinguish empirically among the alternative models for CC subutility $\phi$. 
The SAM model $\phi(y_0, y_{-0}) = y_0 - c\Bar{y} $ satisfies the FOC $\phi_1 =1$ everywhere. Thus every
choice of CC is a best response, so the model makes no testable prediction.
If the model is modified to include a coefficient $c_o$ on the first term, then CC is predicted to be
extreme, either zero or all of earned income $w_0$ as $c_o \lessgtr 1$. 
Again, this is not a useful prediction.

The Conformity model $\phi(y_0, y_{-0}) = -c(y_0 - \Bar{y})^2$
has FOC $1=\phi_1 = -2c(y_0 - \Bar{y})$, yielding the unique best response
$y^* = \Bar{y} - \frac{1}{2c}$, 
truncated below at zero. That is, everyone wants to come in below the mean: just slightly below if $c = c_e$ is large (strong conformity) or far below if $c_e$ is small. 
The comparative statics implication is that, when interior, a player's observed CC will track changes 1:1 in peers' mean CC.
Since not everyone can be below average, things unravel in equilibrium and everyone's CC approaches zero.

For the Rank-dependent model $\phi(y_0, y_{-0}) = C(r)$,
the best response defined by (\ref{eq:brprob}) is an integer programming problem.
Which rank $r$ the player prefers to target depends sensitively on the specific values of $C(1)\geq C(2)\geq ...\geq C(N)$ and of $y_1 \geq y_2 \geq...\geq y_N$.
The key insight is that within the targeted rank interval $(y_r , y_{r-1}]$, the marginal benefit of increasing 
own CC $y_0$ is zero, while marginal cost remains at 1. 
Thus each player will choose CC as low as possible while achieving some particular rank.
Thus
\begin{prediction} (RDM).
According to the Rank-dependent model, positive observed best response CC will only slightly exceed the next ranked choice. 
\end{prediction}
\noindent
Note the contrast to the conformity model prediction, which can accommodate observed CC  near the upper endpoint $y_{r-1}$ of the interval for a specified rank.

It turns out that the predictions of the PEP model 
$\phi(y_0, y_{-0}) = \frac{c_e}{N} \sum_{i=1}^{r-1} [y_0 -y_i ] + \frac{c_p}{N} \sum_{i=r}^N [y_0 -y_i]$
depend sensitively on parameter values: 
\begin{restatable}[PEP]{prediction}{PEPcases}
Let $y^*$ solve the best response problem (\ref{eq:brprob}) for the PEP Model. Then
\begin{enumerate}
\item $y^* = 0$ if $c_e, c_p < 1$,
\item $y^*  = w_o$ if $c_e, c_p >1$ ,
\item  either $y^*= 0$ or $y^* = w_o$ if $c_p > 1 > c_e $, 
depending on the exact values of $c_e , c_p$ and the distribution of peers' choices.
\item if  $c_e > 1 > c_p $ then for any distribution the player targets a particular rank that depends on the exact values of $c_e , c_p$.   
\end{enumerate}
\end{restatable}
\noindent

\noindent
See Appendix \ref{ssec:AppPEPanal} for a proof. 
The underlying intuition is that the marginal benefit $\phi_1$ of increasing CC 
in the PEP model is 
constant within each interval of a specified rank:
it is a weighted average of $c_p$ and $c_e $, with 
weights proportional to the number of upward vs downward comparisons.
This marginal benefit
jumps when the player's CC crosses that of any peer,
as that changes the upward vs downward counts.
In the $c_e > 1 > c_p $ case, the weighted average falls to $c_p<1$ at rank $r=1$,
so the marginal incentive $\phi_1 - 1<0$ is to reduce CC, but it rises to 
$c_e>1$ at rank $r=N+1$ implying an incentive to increase CC. 
At the targeted rank, the marginal incentive is as close to zero as possible. 

We don't observe players' personalized $c_e$ and $c_p$, but if they satisfy $c_e > 1 > c_p $, then the CC target $y^*$ moves in lockstep with adjacent peers' choices $y_r, y_{r-1}$ 
(up to the point where the affordability constraint $y_0 \leq w_0$ binds). 
Therefore a parallel shift in peers'  CC distribution
will provoke an equal shift in $y^*$. 
Perhaps counterintuitively, the PEP model with $c_e > 1 > c_p $ predicts that $y^*$ responds to rank, i.e., to the \emph{number} of upward and downward comparisons, but not to the the magnitude $(y_0 - y_i)$ of those comparisons. 

The analysis is simpler for the MEP model $\phi(y_0,{y}_{-0})= c[y_0-\bar{y}]$,
where $\phi_1 = c= c_p $ when $y_0>\Bar{y}$ or $\phi_1 = c = c_e $ when $y_0<\Bar{y}$. 
As with PEP, the net incentive $\phi_1 -1$ pushes the best response to extremes, $y^* = 0$ or $w_0$, unless $c_e > 1 > c_p \geq 0$. In that case, the net incentive is positive (resp. negative) whenever $y_0<\bar{y}$ (resp. $y_0>\bar{y}$), 
so the best response is $y^* = \Bar{y}$.
The implication is that a player's CC moves in lockstep with peers' mean CC. Assuming convergence towards Nash equilibrium, the prediction is that, over time, dispersion within each peer group will decrease towards zero.

\begin{figure}[!htb]
    \centering
    \begin{tikzpicture}
        \begin{axis}[
            axis lines = middle,
            xlabel = {CC gap $x=(y_0 - y_i)$},
            ylabel = {Gradient $V'(x)$},
            ticks = none,
            xmin = -2, xmax = 2,
            ymin = -2, ymax = 2,
            every axis x label/.style={at={(ticklabel* cs:1)}, anchor=north},
            every axis y label/.style={at={(ticklabel* cs:1)}, anchor=east},
        ]
            \addplot[domain=0:4, blue, ultra thick, samples=100] 
            {0.6*exp(-0.6*x)}; 
            \node[blue] at (axis cs: 1.5, 1.1) {Pride zone};

            \addplot[domain=-4:0, red, ultra thick, samples=100] 
            {1.8*exp(1.8*x)}; 
            \node[red] at (axis cs: -1.5, -1.2) {Envy zone};
        \end{axis}
\end{tikzpicture}
\caption{Gradient of piecewise CARA value function $V$. The limit as $x\searrow 0$ (blue curve) is $c_p$ and as $x\nearrow 0$ (red curve) is $c_e$. 
In the example shown,  $(c_p, c_e)=(0.6, 1.8)$. }    
   \label{fig:PTvalfnGrad}
\end{figure}
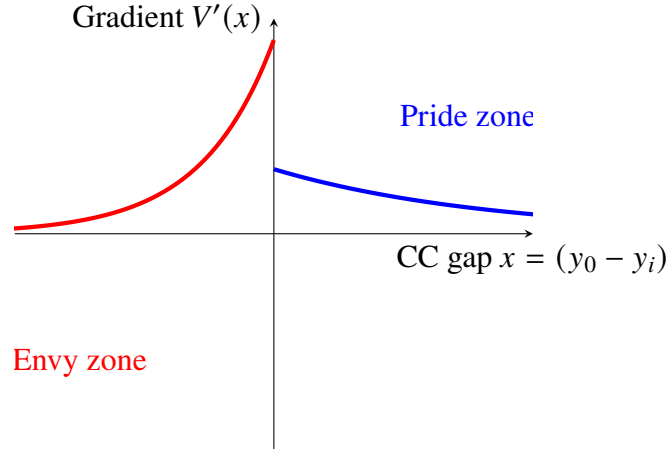

The corresponding Prospect Theory model, PT-MEP, has FOC
$1=V'(y_0 - \bar{y}).$ 
It can be visualized in Figure \ref{fig:PTvalfnGrad}  below as the intersection of a blue or red marginal benefit curve $V'$ with the marginal cost curve, the horizontal line at height 1. 
Intersections $y^*<\bar{y}$ in the Envy zone (red curve) are local minima, not maxima, since the S-shape of $V$ (specifically, its convexity in losses) dictates that $V''(x) >0.$ An intersection
in the Pride zone (blue curve) has $V''(x) <0$ (or concavity in gains) and is a local maxima. 
Indeed, since the blue curve is strictly decreasing from its vertical intercept at $c_p$, such an intersection exists and represents the unique best response when  $c_p>1$. 
In the CARA parametrization, 
the FOC 
$1=V'(x)=c_p e^{-c_p x}$  has solution $x^* = \frac{\ln c_p}{c_p}>0$ or $y^* = \bar{y}+ \frac{\ln c_p}{c_p}$.
That is, every player with $c_p>1$ wants \emph{above}-average CC. 
Players with $c_e \geq 1 \geq c_p $ have best response $y^* = \bar{y}$, and those with 
$c_e, c_p < 1$ have $y^* = 0$. 
Thus we have

\begin{prediction} (PT-MEP).
According to the PT-MEP model, each player will choose either CC level $y^*= 0$, or else a level $y^*\geq \bar{y}$ that increases 1:1 in $\bar{y}$ as long as income permits.
\end{prediction}
\noindent
Note that this prediction is the polar opposite of the conformist model, which says everyone aims for CC somewhat below the mean.

The analysis of the PT-PEP model 
uses similar logic but is more nuanced.
Here the marginal benefit  $\phi_1 =  \frac{1}{N}\sum_{i=1}^N V'(y_0 - y_i)$
can be visualized in 
Figure \ref{fig:PTvalfnGrad} as the unweighted average of the heights of the red and blue lines evaluated at each comparison. 
Appendix \ref{ssec:AppPTPEPanal}
shows that the best response in PT-PEP tracks parallel shifts in peers' CC. 
The analysis also discloses a distinctive implication of 
the S-shaped value function:  
diminishing sensitivity as captured in $V$ means that 
a unit shift in $y_i$ has less impact when it is more remote from $y_0$. 
The upshot is
\begin{restatable}[]{prediction}{PTPEPpred} (PT-PEP).
According to the PT-PEP model, each player either will choose  $y^*= 0$  or else will choose $y^*> 0$ that, as long as income permits, (1) increases 1:1 in $y_{-0}$ 
and (2) decreases in pairwise mean-preserving spreads. 
\end{restatable}

We'll see later that, although other models also predict E(1), they differ from 
the PEP model on E(2).
The intuition is that 
mean-preserving spreads in peers' CC are irrelevant in MEP and PT-MEP, and that they boost CC (not decrease it) in the piecewise linear PEP model to the extent that envy is stronger than pride.



\section{Experiment Design}\label{sec:design}

The experiment was conducted in person between November 2024 and  January 2025 in the ESSEXLab at the 
University of Essex, using oTree programs \citep[][]{chen2016otree} and SONA recruitment software. 
Each of the 14 sessions consisted of 16 human participants drawn from the ESSEXLab subject pool, for a total of 224 participants. 
After checking into the lab, each participant was seated at a workstation and chose a unique username to be used for the entire session. 
To avoid unseemly choices or disclosure of true personal identities, available usernames were computer generated by concatenating a random adjective, a random animal, and a random digit; 
participants (referred to henceforth as players) could extend the list indefinitely to find an acceptable name. 
Players then received written instructions
which were also read out aloud. They took a comprehension quiz which  focused on 
understanding the real effort task and the user interface.

After completing the quiz, players were randomly assigned to a permanent peer group of four, and never interact with anyone outside that group. Each session consisted of a first block of 10 rounds followed by a second block of 5 rounds.  

\begin{figure}[H]
    \centering
    \includegraphics[width=0.5\linewidth]{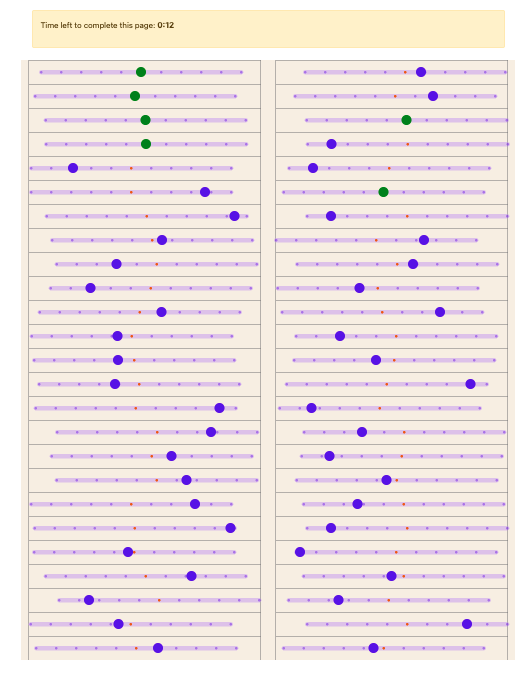}
    \caption{Slider task interface}
    \label{fig:SliderTask}
\end{figure}

\paragraph{First block.}
Each round started with a standard real effort task known as sliders  \citep{gill2012structural}, in which a screen displayed 50 blue sliders at random positions as in Figure \ref{fig:SliderTask}. The player's task was to drag as many sliders as possible to the 50\% mark (middle) within 60 seconds. Once this mark was reached, the slider turned green. For each green slider a player earned 1 token with a cash value 50 pence = 0.5 GBP, about 0.65 US dollars at  then-current exchange rates.

\begin{figure}[H]
    \centering
    \includegraphics[width = .25\textwidth]{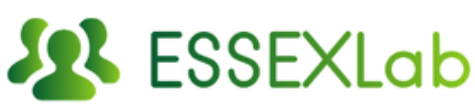}
    \caption{A decal.}
    \label{fig:Decal}
\end{figure}

\begin{figure}[H]
    \centering
    \includegraphics[width=0.5\linewidth]{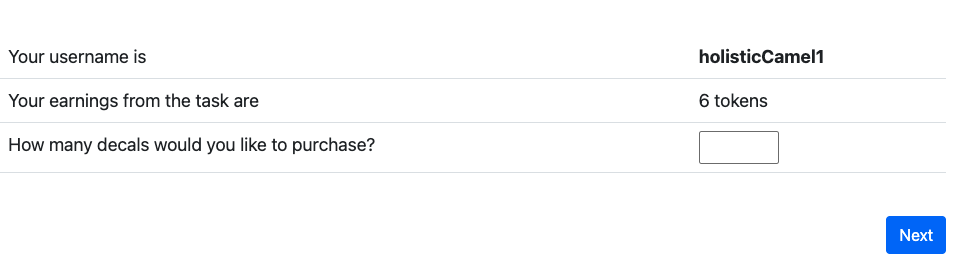}
    \caption{Decal purchase decision screen}
    \label{fig:ConsumptionStage}
\end{figure}

\noindent
At the next stage of each round, players had the opportunity to use earned tokens to purchase indivisible units of a good called ``decals'' at a fixed price of one token per decal.
As illustrated in
Figure \ref{fig:Decal},
the decals feature the ESSEXLab logo; they exist only in digital form and vanish at the end of each round. 
A player who earned $w_0$ tokens in the first stage of the round could purchase any number of decals from zero to $w_0$.
\noindent
Players entered their purchase decisions
in a text box as illustrated in 
Figure \ref{fig:ConsumptionStage}. The box reminded them of their chosen username and their available purchasing power.\footnote{ 
Instructions reminded players that ``Decals have no monetary value and will not affect your final payment. For example, if you earn 21 tokens from Stage 1 and choose to purchase 10 decals,
you will be paid 11 tokens if that round is selected for payment.'' See Appendix \ref{sec:ExperimentalInstructions} for the full text.} 

\noindent
After all four group members had made their purchase decisions, 
they proceeded to a third stage where they were shown how their decal purchase compared to others in the group. The top of the display reminded each trader of their own earned and unspent tokens (see Figure \ref{fig:TreatmentDifferences} and Section \ref{ssec:JvVtreats} for details).
After they had observed the comparison screen, the  group advanced to the next round.

\paragraph{Second block.}
After completing the 10 rounds in block 1, players received new instructions for an additional 5 rounds. 
As before, players earned tokens in the real effort task, made their decal purchase decision and were presented with the summary of choices. In the additional rounds, every player was then offered the unilateral opportunity to  revise their own purchase decision.
Players were told that \textbf{one out of the four} revisions within their group would be chosen randomly to be  implemented while the other three players would keep their initial decal choice.
Thus each player had the opportunity to best-respond to other players' actual choices, and the corresponding choice was realized with probability 0.25.  Again players used the screen illustrated in Figure \ref{fig:ConsumptionStage} to make a revision, and Figure \ref{fig:TreatmentDifferences} illustrates the comparison screen both before and after the realized revision. 

At the end of the second block, players completed a standard questionnaire and the computer randomly selected one of the 15 rounds for payment. All unspent tokens in that round were redeemed using the electronic payment system Tremendous at the rate of 0.5 GBP per token. 
Players' earnings ranged from  \pounds 5 to  \pounds 31.5  with a mean of \pounds 22.58, about US \$30 at the then-current exchange rate, for a bit less than one hour in the lab.

\subsection{Treatments}\label{ssec:JvVtreats}

We employed two treatment variables. The first was within session and was just described: players' moves are either \textbf{Simultaneous} (in the first block, Rounds R1-10) or else they have the \textbf{Last Move} (in the second block, R11-15). 

\begin{figure}[H]
    \centering

    \begin{subfigure}[b]{0.45\textwidth}
        \includegraphics[width=\textwidth]{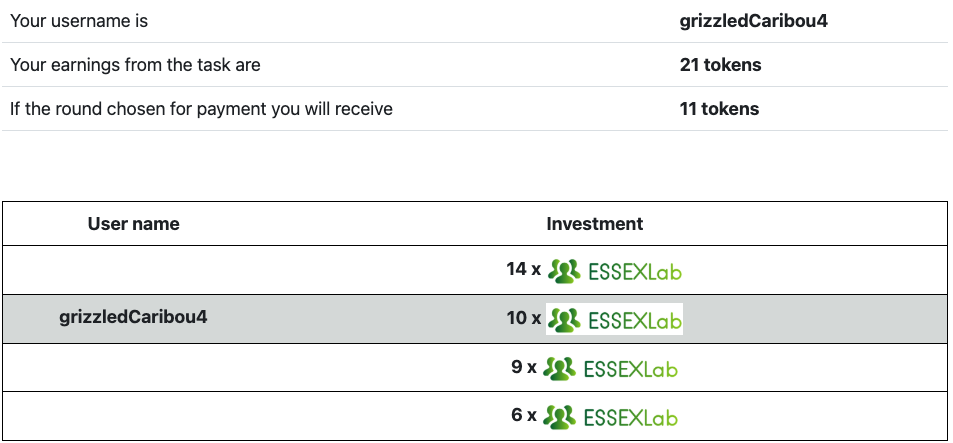}
        \caption{Jones}
        \label{fig:NV}
    \end{subfigure}
    \hfill
    \begin{subfigure}[b]{0.45\textwidth}
        \includegraphics[width=\textwidth]{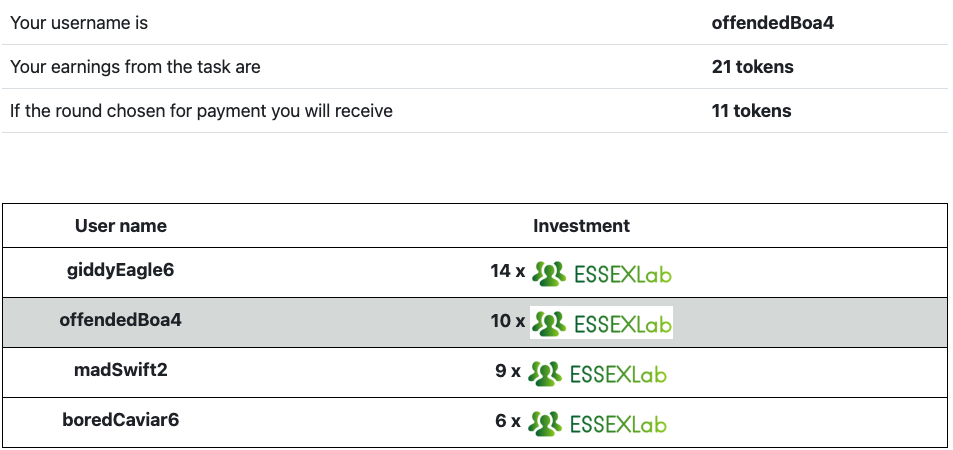}
        \caption{Veblen}
        \label{fig:CV}
    \end{subfigure}
    \hfill

    \caption{Display stage screenshots. }
    \label{fig:TreatmentDifferences}
    
\end{figure}

Our other treatment variable concerns the visibility of conspicuous consumption,
and was randomly sequenced across sessions. 
As in Panel (a) of Figure \ref{fig:TreatmentDifferences},
peers' usernames are suppressed in the seven sessions using the \textbf{Jones} treatment.
In this treatment peers know little about a given player's level of conspicuous consumption, but the given player sees clearly where she fits into the group's distribution. 
In the other seven sessions with the \textbf{Veblen} treatment, players observe peers' usernames. 
Each player knows that her peers see her exact CC expenditure. 
A side effect is that players are reminded each round in Veblen that group composition remains the same across all rounds, so peer group constancy may be more salient than in the Jones treatment.\footnote{We should note, however, that a lead sentence in the instructions for both treatments says, ``You will be assigned to a group of four participants, which will remain the same for the entire experiment.'' }

The idea is that the Veblen treatment provides a platform for social image concerns about how others see me, while 
the Jones treatment drastically reduces the scope for such concerns since my peers there can infer very little about my personal level of conspicuous consumption. 
Our Jones treatment still
provides a platform for self image concerns, inasmuch as it enables me to see exactly where my own conspicuous consumption fits in with peers'.
That contrast captures our interpretation of the main visibility issues discussed in existing literature. 
However, we recognize that those discussions are quite diverse and that other interpretations are possible. 

\subsection{Hypotheses}
Our laboratory data consist of 14 sessions, half of them Veblen and half Jones, each with 4 isolated peer groups. 
Each of those 56 independent trials consists of 4 players' earned income and CC choices over 15 periods, 10 of them Simultaneous and 5 of them Last Move. To conclude this section we list hypotheses  to be tested against those data. 
The first three hypotheses speak to the basic nature of conspicuous consumption but are not linked to specific models.

\begin{hyp}
\label{h:TreatmentDifference}
Conspicuous consumption under the Veblen treatment is higher than under the Jones treatment.
\end{hyp}

\noindent
The main rationale is that, in addition to the self-image concerns enabled in the Jones treatment, the visibility to peers in the Veblen treatment enables social image concerns,
and those additional concerns may boost conspicuous consumption. \\

\begin{hyp}
\label{h:decay}
Conspicuous consumption tends to decrease in later rounds in each block and in both Veblen and Jones treatments.
\end{hyp}
\noindent
This hypothesis is not based on our theory, which is static, but rather on very loose analogy to public goods games, where group-oriented investment tends to decrease over time (see, e.g., \cite{ledyard1995public}). 
Actually, we hope that H2 will turn out to be false, and that CC will stabilize in some treatments.\\

\begin{hyp}
\label{h:Jump}
Conspicuous consumption in the second block is higher than in the first block.
\end{hyp}

\noindent
The second block offers the opportunity to unilaterally adjust conspicuous consumption. This  may make conspicuous consumption more attractive because it eliminates strategic uncertainty about the distribution of peers' conspicuous consumption. 
An extra advantageous adjustment opportunity also might make the game more engaging. \\

The remaining hypotheses are based on the Predictions regarding best responses developed in Section \ref{ssec:ModAnal} and the surrounding discussion.
They therefore apply to the post-adjustment choices in Rounds R11-15.


\begin{hyp}
\label{h:QL}
Conspicuous consumption is not correlated with income $w_o$.
\end{hyp}
\noindent
This comes directly from Prediction A, which reminds us that income effects are largely absent from our quasi-linear models. 
It is best tested on the subset of choices that are interior, omitting the corner choices of zero CC (which would mechanically tend to confirm the hypothesis) and of 100\% CC (which are not covered by Prediction A).
It is conceivable that earned income is correlated with peers' consumption choices, so it would be desirable when testing this hypothesis to control for the mean $\bar{y}$ of
those choices.\\

\begin{hyp}
\label{h:OridnalModel}
Positive CC choices are targeted: they will lie just above a peer's choice. 
\end{hyp}
\noindent
As elaborated in Prediction B, this is the distinctive implication of the Rank-dependent model, in which players 
spend the minimal amount of tokens to achieve a desired rank.\\

\begin{hyp}
\label{h:MEP}
Conspicuous consumption is correlated with the mean of peers' CC.
\end{hyp}
\noindent
As noted in Predictions C - E and in the surrounding discussion, Hypothesis \ref{h:MEP} follows from most of the models we consider. 
Except for SAM and the Rank-dependent model, the prediction is that own CC $y^*$ moves 1:1 in response to parallel shifts in peers' CC. 
Peers' mean CC $\bar{y}$ serves as a rough proxy for parallel shifts for all relevant models, and is an exact proxy for MEP and PT-MEP. 
The
Conformity model further predicts that $\bar{y}$ will move towards 0 in later periods.

The last hypothesis concerns a more diagnostic statistic.
We
decompose peers' mean CC $\bar{y}$ into upward comparisons $L$ to ``leaders'' with greater CC and downward comparisons $D$ to peers with lower CC than the focal agent with rank $r$, so 
\begin{equation}\label{eq:Ldef}
\bar{y} \equiv \frac{1}{N}\sum_{i=1}^{N} y_i
= \frac{1}{N}\sum_{i=1}^{r-1} y_i + \frac{1}{N}\sum_{i=r}^{N} y_i \equiv L + D 
\end{equation}
Appendix \ref{ssec:AppPTPEPanal}
shows that a pairwise mean-preserving spread in peers' CC will increase $L$ and decrease $y^*$. 
Thus, holding $\bar{y}$ constant, $L$ is a convenient proxy for dispersion of peers' CC.
\begin{hyp}
\label{h:PEP}
Controlling for $\bar{y}$, conspicuous consumption is negatively correlated with 
the variable $L$.
\end{hyp}

\noindent
This comes directly from Prediction E for the PT-PEP model.
By contrast, for reasons noted earlier, MEP and PT-MEP predict zero correlation, and 
PEP predicts positive correlation with L.

\section{Results}\label{sec:results}
We begin with an overview of the data, using summary graphs and tables.
Later subsections will report hypothesis tests and call out the main results.

\begin{figure}[H]
    \centering

    \begin{subfigure}[t]{0.49\textwidth}
        \centering
        \includegraphics[width=\linewidth]{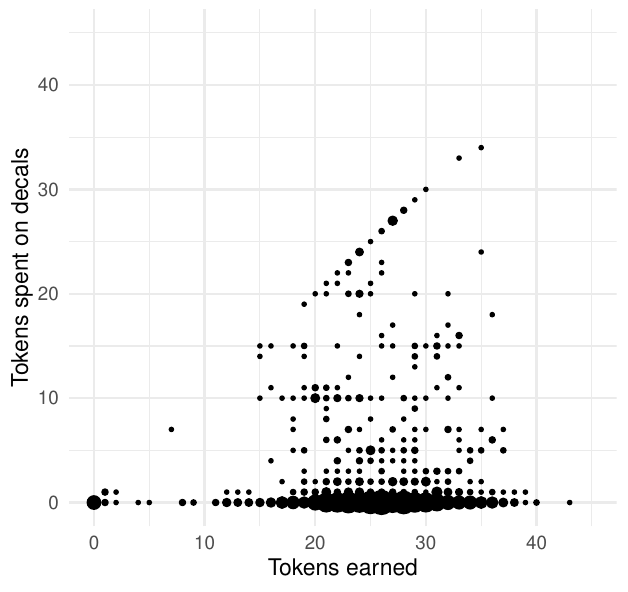}
    \end{subfigure}
    \hfill
    \begin{subfigure}[t]{0.49\textwidth}
        \centering
        \includegraphics[width=\linewidth]{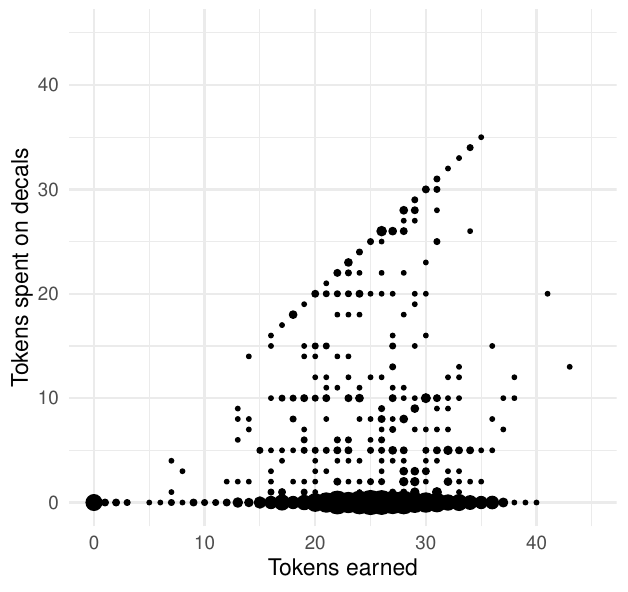}
    \end{subfigure}

    \vspace{0.3em}

    \begin{subfigure}{0.20\textwidth}
        \centering
        \includegraphics[width=\linewidth]{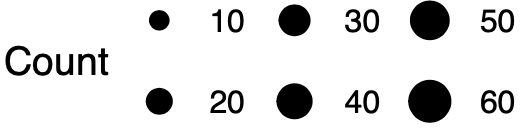}
    \end{subfigure}

    \vspace{0.3em}

    \caption{Scatterplot of tokens earned $w_0$  and tokens spent $y_0$ on conspicuous consumption by all players in all Simultaneous Move Rounds (R1-10). Jones (left) vs Veblen (right).
    }
    \label{fig:Scatter_Earnings_ConspCons}
\end{figure}

\subsection{Conspicuous Consumption in Rounds 1-10}
Figure \ref{fig:Scatter_Earnings_ConspCons} shows tokens earned in the real effort task ($w_0$) and tokens spent on decals ($y_0$) 
by individual players, pooling across Simultaneous Move Rounds (R1-10). 
Below, we will usually refer decal expenditures as observed conspicuous consumption (CC).
The medium size dots near the origin of both graphs indicate that some
players ignore the task in some rounds.
The clusters of large dots along the horizontal axis indicate that many players do not engage in CC.
The dots on the main diagonal indicate that some players spend all their earnings on CC in some rounds, while dots between horizontal axis and diagonal represent interior choices.
The distributions for Jones (left) and Veblen (right) treatments seem roughly similar, with perhaps a few more points on the diagonal for Veblen.

\begin{figure}[H]
    \centering
\begin{subfigure}[t]{0.49\textwidth}
\includegraphics[width=1\linewidth]{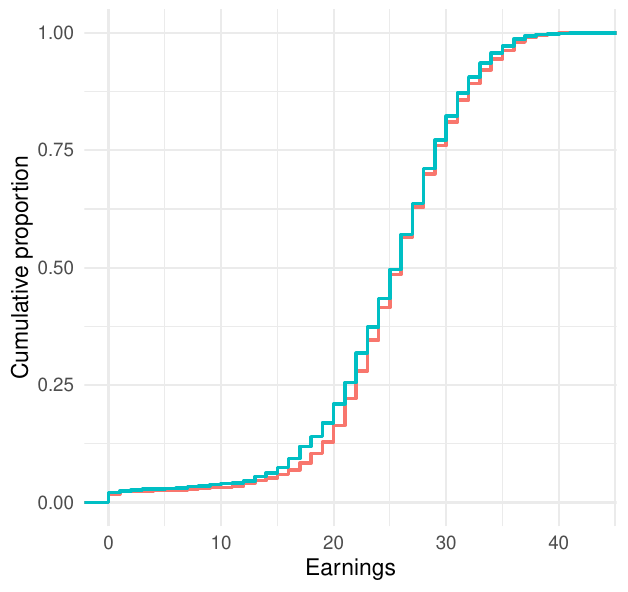}
\end{subfigure}
\begin{subfigure}[t]{0.49\textwidth}
    \includegraphics[width=1\linewidth]{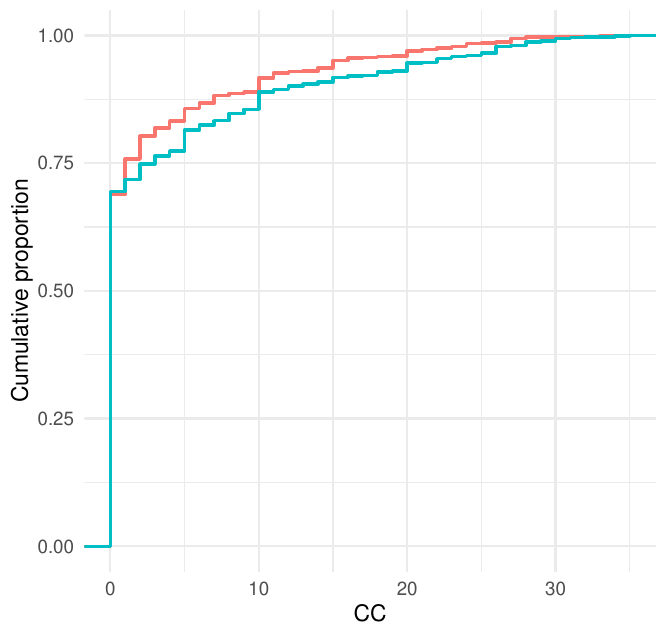}
\end{subfigure}

\vspace{0.3em}

\begin{subfigure}{0.28\textwidth}
\centering
\includegraphics[width=\linewidth]{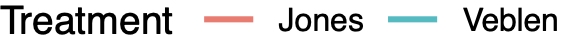}
\end{subfigure}



\caption{Cumulative Distribution Functions (cdf's) for earnings (left) and conspicuous consumption CC (right) in R1-10. 
    }
    \label{fig:cdfs4ConspConsEarnings}
\end{figure}

Figure \ref{fig:cdfs4ConspConsEarnings} collapses the bivariate distributions in Figure \ref{fig:Scatter_Earnings_ConspCons} into 
univariate distributions for earnings and CC expenditures.
The left panel shows that most players earn between 20 and 30 tokens; the
slight rightward shift of the Jones (red)  line  suggests perhaps slightly greater earnings  relative to Veblen (blue). 
The right panel indicates that almost 2/3 of players in an average round
choose zero CC,
but among those with positive CC, expenditures tend to be larger in the Veblen treatment.

\begin{figure}[H]
    \centering
\begin{subfigure}[t]{0.99\textwidth}
    \includegraphics[width=1\linewidth]{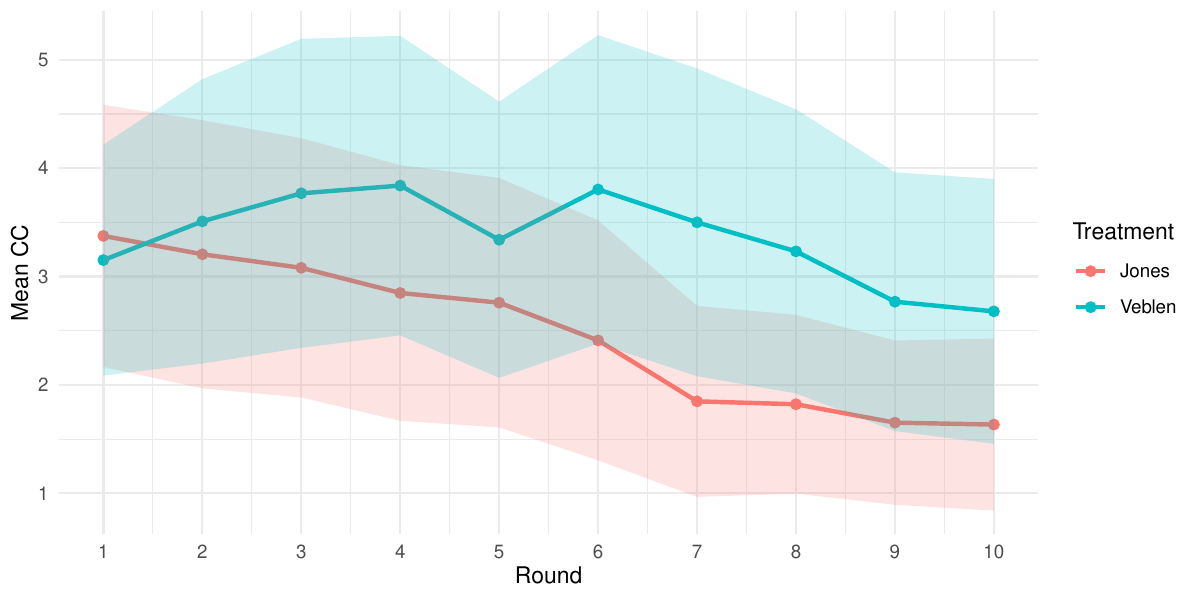}
\end{subfigure}
\begin{subfigure}[t]{0.48\textwidth}
        \includegraphics[width=1\linewidth]{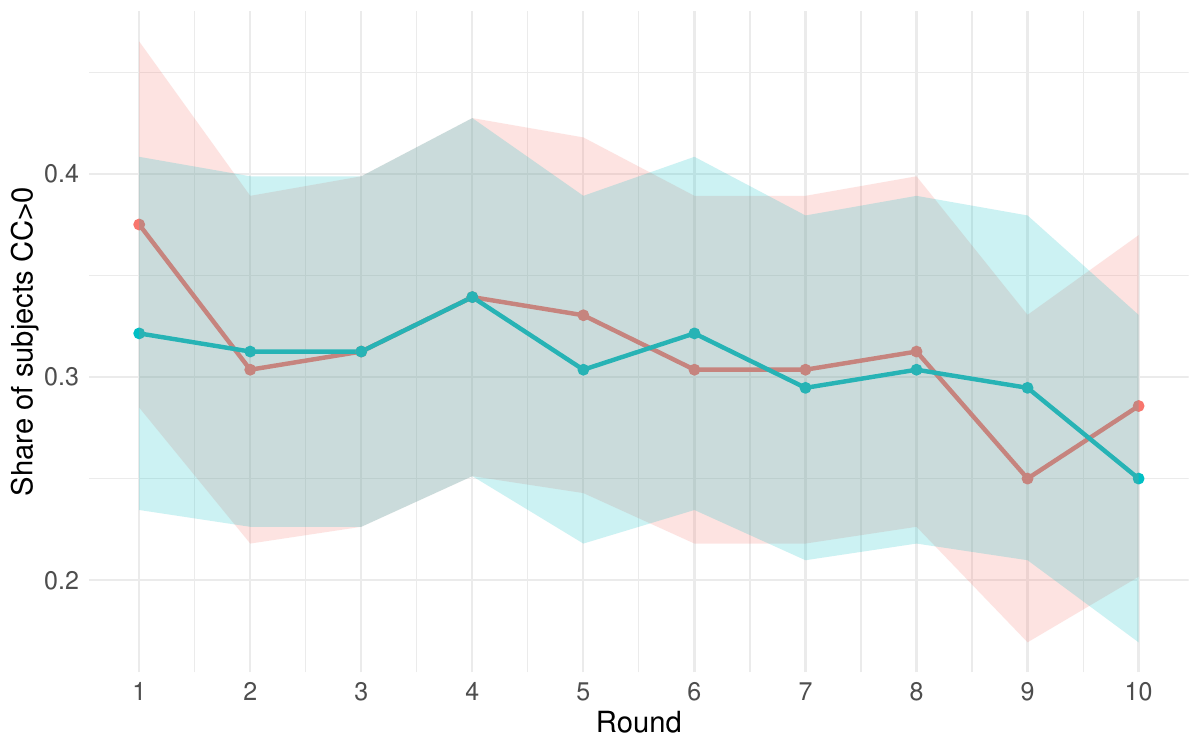}
\end{subfigure}
\begin{subfigure}[t]{0.48\textwidth}
        \includegraphics[width=1\linewidth]{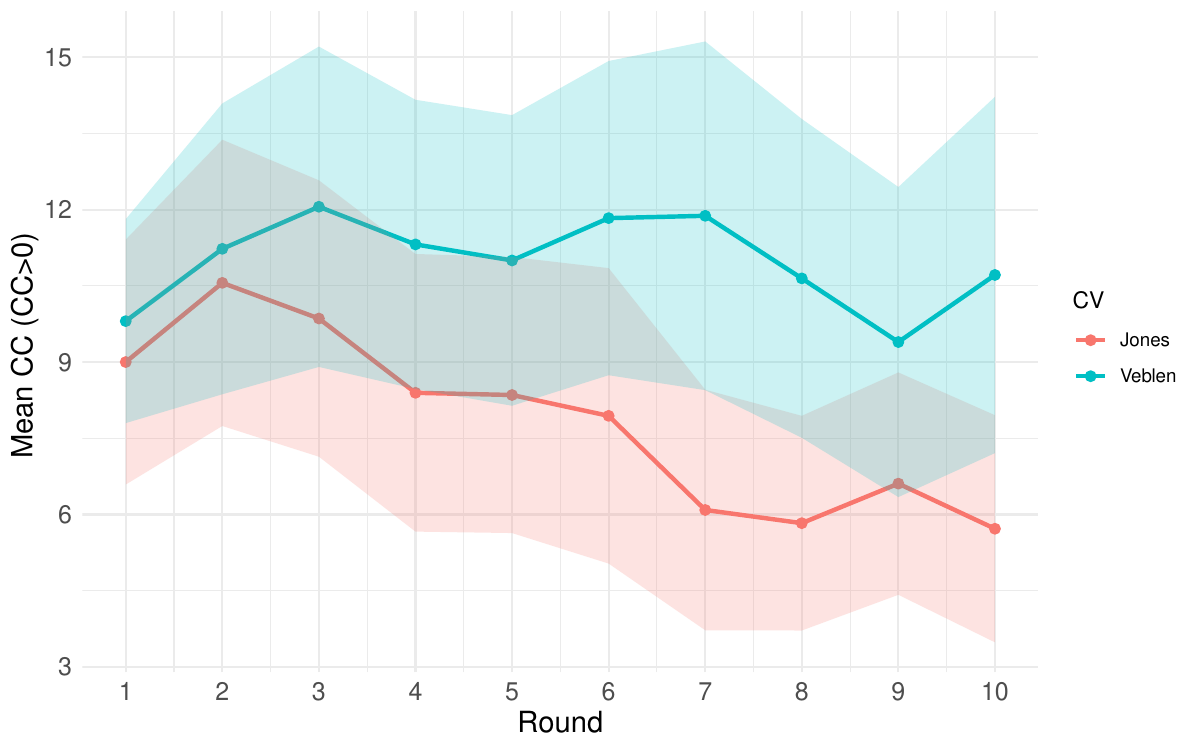}
\end{subfigure}
\caption{Mean conspicuous consumption 
(and 95\% confidence intervals) in 
Rounds 1-10. 
The top panel 
shows overall trends.  
The bottom panels show the share of players with positive CC (left), and mean CC among players whose spending is positive (right).}
    \label{fig:TimeSeries}
\end{figure}

Figure \ref{fig:TimeSeries} shows CC time trends in Rounds 1-10. 
The top panel 
indicates a modest downtrend from around 3.5 tokens per capita in Round 1 to under 3.0 in the Veblen treatment (and to under 2.0 in the Jones treatment) by Round 10.
The bottom left panel
shows that the extensive margin contributes to both downtrends: the fraction of 
players who buy any tokens declines from about 35\% in Round 1 to under 30\% in Round 10 in both  treatments.
The bottom right panel 
focuses on the intensive margin and
shows that, among those with positive CC, per capita spending holds steady at 9-12 tokens in the Veblen treatment, but it trends downward in the Jones treatment from around 9-10 tokens in  Rounds 1 to 3 to 6-7 tokens in Rounds 8 to 10.

To test these impressions more formally, we regress $y_0$, the number of tokens spent on CC,
on earnings $w_0$ and treatment dummy variables,
using OLS with errors clustered at the peer group level.
Table \ref{tab:Regression_VeblenEffect} reports the results.
It confirms that CC is lower by about one token on average in the later rounds (R6-10),
a highly significant difference from earlier rounds.
It also confirms that the Veblen treatment encourages about an extra token of CC on overall, 
but that is not a statistically significant difference given the high degree of variability and modest size of our sample. 
The second column includes an interaction term,
significantly positive at the conventional 5\% level, which indicates a lesser reduction in CC in later rounds of
the Veblen treatment.
Restricting to the subsample of observations with strictly positive conspicuous consumption (CC) choices (column 3) and regressing inclusion in the subsample (column 4) yields similar patterns. The 
last two rows report an F-test on the overall effect of later rounds in Veblen. The p-values indicate that the reduction in CC in Veblen is insignificant at the intensive margin and overall, though borderline significant at the extensive margin ($p<0.10$). 
Appendix Table \ref{tab:Regression_VeblenEffect2} shows similar patterns when the subsample includes only players with conspicuous consumption greater than three tokens.


\begin{table}
\begin{center}
\begin{tabular}{l c c c c}
\hline
 & CC  & CC  &  Only CC$>0$ & $I_{CC>0}$\\
\hline
(Intercept)                  & $0.76$        & $0.97$        & $5.88^{*}$    & $19.16^{*}$   \\
                             & $(0.90)$      & $(0.93)$      & $(2.59)$      & $(7.69)$     \\
earnings          & $0.09^{*}$    & $0.09^{*}$    & $0.14$        & $0.59^{*}$   \\
                             & $(0.04)$      & $(0.04)$      & $(0.10)$      & $(0.29)$     \\
Veblen                       & $0.94$        & $0.53$        & $1.88$        & $-0.98$      \\
                             & $(0.76)$      & $(0.82)$      & $(1.61)$      & $(5.85)$     \\
R6-10        & $-1.02^{***}$ & $-1.43^{***}$ & $-3.14^{***}$ & $-5.78^{**}$ \\
                             & $(0.25)$      & $(0.31)$      & $(0.87)$      & $(1.78)$     \\
Veblen $\times$ R6-10 &               & $0.82^{*}$    & $2.68^{*}$    & $1.34$       \\
                             &               & $(0.38)$      & $(1.09)$      & $(2.40)$     \\
\hline
Num. obs.                    & $2240$        & $2240$        & $691$         & $2240$       \\
F-test: Veblen R6-10 effect = 0 & n/a          & F = 2.449    & F = 0.275    & F = 2.892    \\
p-value                         & n/a          & 0.118        & 0.600        & 0.089        \\
\hline
\multicolumn{5}{l}{\scriptsize{$^{***}p<0.001$; $^{**}p<0.01$; $^{*}p<0.05$; $^{\circ}p<0.1$}}
\end{tabular}
\caption{Regression coefficient estimates 
(and standard errors clustered at group level)
for conspicuous consumption (columns 1 and 2), for CC in restricted subsample where it is  positive (column 3), and for Indicator dummy [$I_{CC>0}$=100 if CC is positive and =0 ow] (column 4). Baseline explanatory variables are early rounds (R1-5) and Jones treatment.   }
\label{tab:Regression_VeblenEffect}
\end{center}
\end{table}

Combining these regression results with the earlier descriptive statistics, we can draw two conclusions about conspicuous consumption in Rounds 1-10. 

\noindent
\textbf{Result 1.} Consistent with Hypothesis \ref{h:TreatmentDifference},
conspicuous consumption is enhanced in the Veblen treatment, especially so in later rounds of Block 1.   

\noindent
\textbf{Result 2.} Consistent with Hypothesis \ref{h:decay}, conspicuous consumption overall tends to be lower in later rounds of Block 1, by about 1 token per capita from a baseline of about 3 tokens. Much of the decline comes at the extensive margin: a smaller fraction of players purchase any decals in later rounds. 

The positive interaction terms between treatment and Rounds 6-10 dummies suggest that the Veblen treatment better sustains CC over time, primarily through an increase in CC intensity.

\subsection{Last Move treatment}
\begin{figure}[H]
    \centering
\begin{subfigure}[t]{0.9\textwidth}
    \includegraphics[width=1\linewidth]{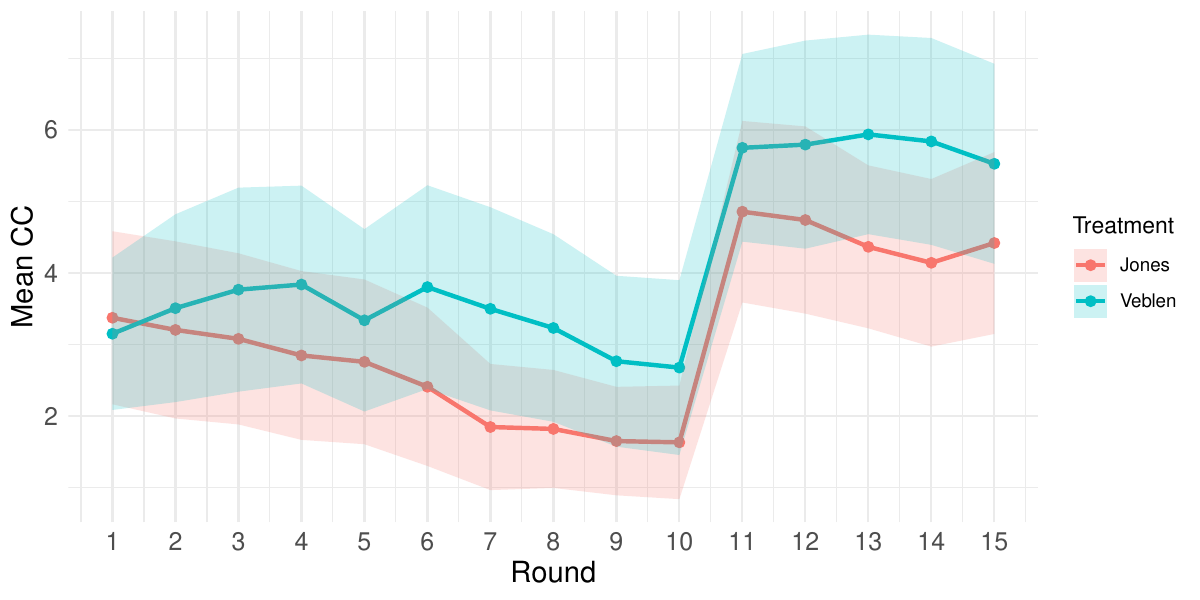}
\end{subfigure}
\begin{subfigure}[t]{0.45\textwidth}
        \includegraphics[width=1\linewidth]{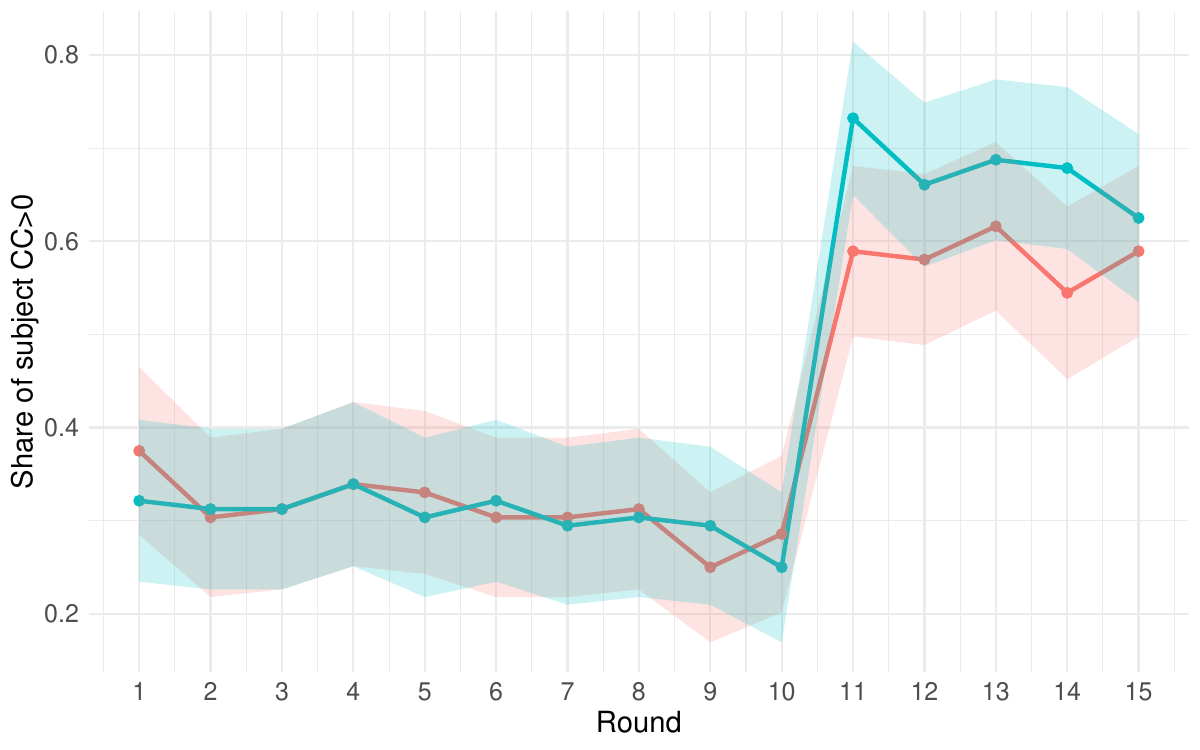}
\end{subfigure}
\begin{subfigure}[t]{0.45\textwidth}
        \includegraphics[width=1\linewidth]{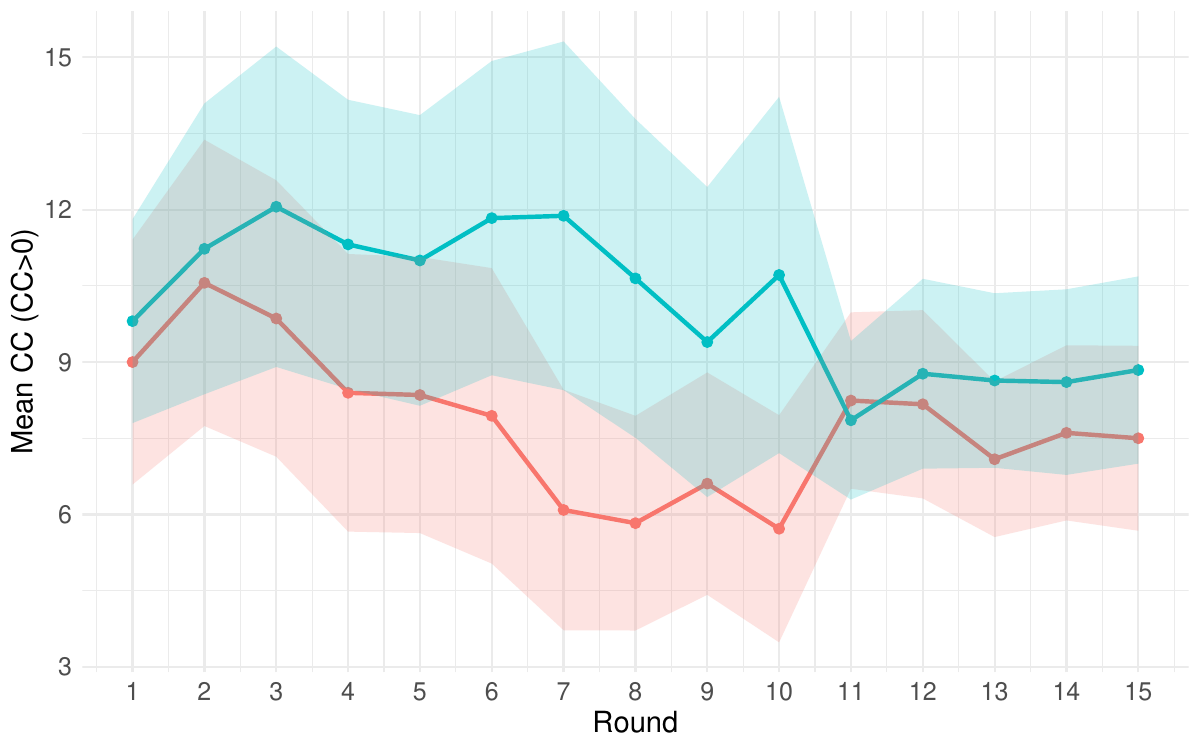}
\end{subfigure}
    \caption{Mean conspicuous consumption 
(and 95\% confidence intervals) in all rounds. The top panel shows overall trends. The bottom panels show the share of players
with positive CC (left), and mean CC when player’s expenditure is positive (right).} 
    \label{fig:TimeSeries2}
\end{figure}

We now consider what happens in Rounds 11 - 15, when we allow players to adjust their CC as last movers in response to their peers' CC. 
Panel (a) of Figure \ref{fig:TimeSeries2} shows a dramatic and immediate impact:
per capita CC almost doubles in Period 11 from its levels in period 10, and stays elevated through period 15. 
Panel (b) shows that the jump is almost entirely at the extensive margin: the fraction of players choose to consume any CC roughly doubles.
Panel (c) indicates that, among those players, per capita CC generally remains higher in Veblen than in Jones, with little apparent time trend. 

\begin{table}[H]
\begin{center}
\begin{tabular}{l c c c}
\toprule
 & CC & $I_{CC>0}$ & Only CC$>0$\\
\midrule
(Intercept)                   & $2.46^{***}$ & $31.16^{***}$ & $7.91^{***}$    \\
                        & $(0.71)$     & $(7.15)$      & $(1.14)$       \\
R11-15        & $2.04^{*}$  & $27.23^{***}$ & $-0.19$         \\
             & $(0.97)$     & $(7.83)$      & $(1.26)$       \\
Veblen                        & $0.90$       & $-0.63$       & $3.09^{\circ}$  \\
              & $(1.17)$     & $(10.24)$     & $(1.68)$       \\
R11-15$\times$Veblen & $0.37$       & $9.91$        & $-2.29$ \\
               & $(1.32)$     & $(11.13)$     & $(1.61)$       \\
\midrule
Num. obs.                     & $3360$       & $3360$          & $914$         \\
\bottomrule
\multicolumn{4}{l}{\scriptsize{$^{***}p<0.001$; $^{**}p<0.01$; $^{*}p<0.05$; $^{\circ}p<0.1$}}
\end{tabular}
\caption{Regression results including Last Move data (R11-15) as well as R1-10.
Dependent variables are conspicuous consumption (CC, col 1), dummy = 100 if CC is positive (col. 2), and CC in positive subsample (col. 3). 
Errors clustered at the peer group level.}
\label{tab:RDD}
\end{center}
\end{table}

Table \ref{tab:RDD} reports a discontinuity regression that confirms significance. The first column shows that overall per capita CC jumps by a significant 2.04 tokens when the Last Move treatment is in effect.
The second column shows that, at the extensive margin,
the share of players with positive conspicuous consumption jumps by a highly significant 27 percentage points. 
The third column shows no
significant action at the intensive margin. The last two rows indicate
that CC tends to be insignificantly larger in the Veblen treatments.
Appendix Table \ref{tab:RDD_invest3} reports similar results when the subsample criterion $CC>0$ is replaced by $CC>3$ for ``serious'' CC,
and Table \ref{tab:InitVsFinalCC} provides a more detailed breakdown of pre- and post-adjustment data.
To reiterate the key point, 

\vspace{.07in}
\noindent
\textbf{Result 3.} Consistent with Hypothesis \ref{h:Jump}, the switch to the Last Move treatment in Period 11 has a large positive impact on conspicuous consumption.
The impact is immediate and is sustained through the last round. It comes at the extensive margin: a larger fraction of players in both treatments engage in CC.\\


\subsection{Model Testing}\label{ssec:modelTesting}

A first step in testing our models is to check whether our quasi-linear framework is adequate. 
Its hallmark, as noted earlier,
is the absence of income effects.
Table \ref{tab:EnvyPrideRegression}, presented below, reports a regression that includes $w_o$ as an explanatory variable for observed CC.
From it we will obtain\\
\noindent
\textbf{Result 4.} Consistent with Hypothesis \ref{h:QL}, but inconsistent with most multiplicative interaction models, the coefficient estimates for income $w_0$ in equation (\ref{eq:MainRegression})
are close to zero and statistically insignificant.\\

\begin{table}[H]
    \centering
    \begin{tabular}{l|c|cc|cccc}
     Treatment    
     &  N & $W_I = 0$ & $W_F = 0$
     & $W_F = W_I$ & $W_F > W_I$ & $W_F < W_I$
     & $q^* $
     \\
     \midrule
     Jones   
     & 184  
     & 57   &  57
     & 164  & 11  & 9
     & 0.347 
     \\ 
     Veblen    
     & 234 
     & 79 &  76 
     & 206 &  17 & 11
     & 0.435
     \\ 
      \midrule
     All    
     & 418  & 136   &  141 
     & 370 & 28  &  20
     & 0.455
     \\ 
     \bottomrule
    \end{tabular}
\caption{Ordinal Waste Counts in Rounds 11-15. 
$N$ is the number of observations that have positive final CC and unambiguous rank (so excluding tied initial CC).
Waste
$W_I$ (resp. $W_F$) is the difference between initial pre-adjustment (resp. final post-adjustment) CC and that of next lower peer, minus 1.
Entries in columns 2-6 are numbers of observations with positive final CC that satisfy the given conditions. 
Last column gives critical error rate $q^*$ for waste-reducing adjustments at which a binomial test rejects the hypothesis $q \leq q^*$ at $p=0.05$ given observations in previous two columns.
}
    \label{tab:OrdinalTheories}
\end{table}

We now test the distinctive prediction of 
the Rank-dependent model: 
that players will achieve a targeted rank at minimal cost.
That is,
the player will either target rank $r=4$ by choosing post-adjustment CC $y^*=0$
or else will target $r=1,2$ or 3 by choosing
$y^*=y_r + 1$, i.e.,  1 token more
than the next lower peer. 
Define ordinal waste $W$ as expenditure in excess of that amount.
Table \ref{tab:OrdinalTheories} shows that,
contrary to the Rank-dependent model,
$W$ is not typically zero.
Column 2 shows that, perhaps unsurprisingly,
pre-adjustment $W$ is zero in only about a third of instances in each treatment; 
the fact that this fraction is no higher
post-adjustment (see Column 3) is difficult to reconcile with the Rank-dependent model. 

Column 4 suggests considerable \emph{behavioral inertia}: in the majority of observations,
ordinal waste is the same in final and initial CC choices, typically because of zero adjustment.
(In a handful of exceptions, 
the final rank differs from the initial rank but the amount of ordinal waste is the same.)

The last three columns show that active adjustment does not tend to decrease ordinal waste. In fact, ordinal waste increases more often than it decreases.
As detailed in Appendix Section \ref{ssec:bintest}, the last column shows that we can confidently reject (at the conventional $p= 0.05$ level) even
the very weak hypothesis that adjustments are intended to reduce wasteful targeting but directional errors occur no more than 35-45\% of the time. 
In sum, we have\\

\noindent
\textbf{Result 5.} Contrary to Hypothesis \ref{h:OridnalModel},
our participants generally do not target a particular rank cost-effectively.\\

\begin{table}[H]
\begin{center}
\begin{tabular}{l c c c}
\hline
 & All & Veblen & Jones \\
\hline
$w_0$ & $-0.02$       & $-0.09$       & $0.06$        \\
      & $(0.07)$      & $(0.09)$      & $(0.10)$      \\
$\bar y$ & $1.44^{***}$  & $1.27^{***}$  & $2.19^{***}$  \\
        & $(0.20)$      & $(0.19)$      & $(0.42)$      \\
$L$    & $-1.53^{***}$ & $-1.17^{***}$ & $-2.22^{***}$ \\
       & $(0.25)$      & $(0.22)$      & $(0.33)$      \\
\hline
Num. obs.  & $410$         & $225$         & $185$         \\
\hline
\end{tabular}
\caption{Coefficient estimates for equation (\ref{eq:MainRegression}). Data are all interior post-adjustment CC ($y^* \in (0,w_0 )$) choices in R11-15. Regression uses   
round fixed effects and clusters errors at group level. $^{***}p<0.001$; $^{**}p<0.01$; $^{*}p<0.05$; $^{\circ}p<0.1$.
}
\label{tab:EnvyPrideRegression}
\end{center}
\end{table}

We now test key implications of the other models, using the regression
\begin{equation}\label{eq:MainRegression}
    y^* = \beta_0 + \beta_1 w_0 + \beta_2 \bar{y} + \beta_3 L + e.
\end{equation}
The dependent variable $y^*$ is post-adjustment CC in the Last Move treatment, rounds R11-15.
Recall that $ w_0$ is a player's real effort earnings in a given round, while $\bar{y} = \frac{1}{3}\sum_{i=1}^3 y_i$ is mean peer CC and 
$L = \frac{1}{3}\sum_{i=1}^{r-1} y_i$ captures shortfalls from peers' CC, a proxy for dispersion. 
Note that all three explanatory variables are pre-determined: $w_0$ is earned at the first stage of the round and $\bar{y}$ is defined at the pre-adjustment stage. 
So is $L$, which is defined with respect to the focal player's pre-adjustment rank.
For reasons discussed earlier, we exclude corner choices $y^* = 0$ and $y^*=w_0$. 

Table \ref{tab:EnvyPrideRegression} reports income coefficient estimates close to zero, as already noted in Result 4 above. 
Estimates for  $\bar{y}$ are highly significantly positive; point estimates here are above 1.0.  
The Appendix reports several variants of this regression, e.g. for subsample of serious CC $(y^*>3)$ and different proxies for dispersion. Those estimated $\bar{y}$ coefficients are all closer to 1.0.
Thus

\vspace{.05in}
\noindent
\textbf{Result 6.} Consistent with Hypothesis \ref{h:MEP} and with many of the models, our players' CC choices on average respond roughly 1:1 to changes in peers' mean CC. \\

More diagnostic is the coefficient estimate for $L$ in (\ref{eq:MainRegression}),
which captures the impact of a mean-preserving spread in peers' CC.
The Conformity, MEP and PT-MEP models predict a zero coefficient, while PEP predicts a positive coefficient.
Inconsistent with those models, Table \ref{tab:EnvyPrideRegression} reports
highly significantly negative estimates for $L$ overall and for each treatment separately.  Appendix
Tables \ref{tab:EnvyPrideRegressionCombined} - \ref{table:active-std} report the results of several robustness checks.
In all variants we have examined, 
when dispersion (however measured) has a significant impact, it is always negative. 

\vspace{.05in}
\noindent
\textbf{Result 7.} 
Consistent with Hypothesis \ref{h:PEP} and the PT-PEP model, but inconsistent with other models we consider, players' CC choices tend on average respond negatively to increased dispersion in peers' CC.\\

Of the models considered, 
the PT-PEP seems most consistent with 
qualitative features of our data. 
The Simple Average Model (SAM) fails to make useful predictions. 
Conformity incorrectly 
(see Appendix Figure \ref{fig:StdDevTrends}) predicts shrinking mean CC and shrinking deviations from the mean.
The Rank-dependent model incorrectly predicts cost-effective rank targeting. 
Peerwise Envy/Pride (PEP) incorrectly predicts a positive impact for L, and Mean Envy/Pride (MEP) and PT-MEP incorrectly predict no impact.

Of course, it is possible to construct other models of conspicuous consumption that deliver a positive impact of the mean and a negative impact of dispersion in peers' CC.
Our attempts to do so, however, all
contradicted some features of our own data and/or the stylized fact that envy is at least as powerful as pride.

\section{Conclusion}\label{sec:conclude}
Our paper is motivated by the overabundance of models of conspicuous consumption (CC). 
Some models ascribe CC to the desire to \emph{Keep up with the Joneses}, 
i.e., to find an appropriate place
in the distribution of peers' CC.
Other models in the spirit of \cite{veblen1899thetheory} ascribe CC to how a player wishes to be perceived by peers.
Either way, existing models differ greatly on 
how a player will respond to peers' CC.

We report a laboratory experiment intended to help sort things out. 
The experiment compares two visibility treatments, one of them relatively anonymous and intended to capture Jones (self-image)  motivations, 
and the other more 
public and intended to capture Veblen (social image) motivations. 
The experiment also compares a Simultaneous Move protocol with a Last Move protocol that enables players to best respond. 
We work out testable implications for a variety of theoretical models, and test them on the data. 

We find that a substantial fraction of our players engage in CC even in our Jones (limited visibility) treatment, and that
our Veblen (public visibility) treatment enhances the intensity of CC and helps sustain it over time. 
The Last Move treatment sharply increases the fraction of players who seriously engage in CC. 

Consistent with our quasi-linear theoretical framework,  income effects are negligible in our data. 
Several of the quasi-linear models we consider correctly predict that a player's CC tends to shift roughly 1:1 with peers' mean CC, but only one of them correctly predicts the generally negative impact we see of dispersion in peers' CC. 

That model, Prospect Theory - Pairwise Envy-Pride or PT-PEP, has three primary features. 
First, \textit{loss aversion} implies that nearby upward comparisons (envy) exert stronger effects than comparable downward comparisons (pride), implying incentives to avoid falling behind. 
Second, \textit{pairwise comparisons} allow individuals to respond to the full distribution of peers’ behavior rather than to a single aggregate statistic such as peers' mean. 
Third, \textit{diminishing sensitivity}, or decreasing marginal response to larger shortfalls or excesses in CC,  stabilizes interior choices, in contrast to linear models like MEP or PEP that tend to predict extreme behavior.
In a nutshell, our results
suggest that conspicuous consumption may arise from reference-dependent loss-averse social comparisons applied at the interpersonal level, rather than from, e.g., conformity or pure rank competition.

Several caveats are in order. Our data are decal purchases, a rather bland form of conspicuous consumption, using income earned in an unexciting slider placement task. 
It is possible (see \cite{zenoupeer} for evidence in a different context) that a different task or different form of CC might produce different results.
In particular, we were surprised by the powerful immediate impact of our Last Move treatment on the extensive margin of CC.
Perhaps some that some of that impact is a restart effect (but apparently not all --- the share of positive CC in our last round remains far higher than in the first round).
 
More generally, replication in different settings, with differently worded instructions, is required to confirm (or to qualify) the robustness of our results.
Such future work might shed more light on  behavioral inertia.
Ideally it would produce enough 
active adjustment data to better estimate the impacts of peers' mean and dispersion,
or even enable 
structural estimation of $c_e$ and $c_p$. 

Future empirical work could also investigate the effect of group size, including the absence of visible peer consumption
(size 1).
Future researchers might use a task that signals intelligence or knowledge, or a more dramatic sort of conspicuous consumption.
On the flip side, they might consider treatments that
do not signal anything, e.g., everyone gets the same income ($w_o =30$, say) after dragging sliders, 
or gets no-effort endowments with the same heterogeneity as in our sessions.


\clearpage

\appendix
\begin{center}
 \huge{\textbf{Appendices}}   
\end{center}

\section{Supplementary data analysis}
\begin{figure}[H]
    \centering

    \begin{subfigure}[t]{0.49\textwidth}
        \centering
        \includegraphics[width=\linewidth]{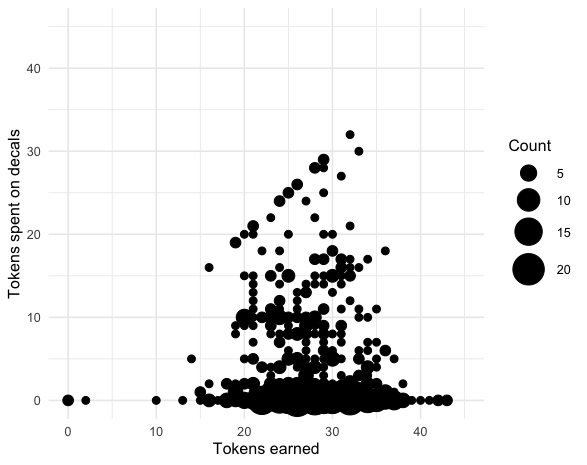}
    \end{subfigure}
    \hfill
    \begin{subfigure}[t]{0.49\textwidth}
        \centering
        \includegraphics[width=\linewidth]{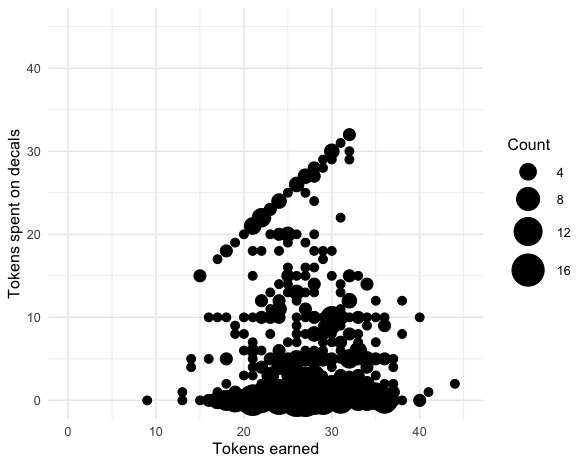}
    \end{subfigure}

    \vspace{0.3em}

    \begin{subfigure}{0.20\textwidth}
        \centering
        \includegraphics[width=\linewidth]{pictures/scatterlegend.jpg}
    \end{subfigure}

    \vspace{0.3em}

    \caption{Scatterplot of tokens earned $w_0$  and tokens spent $y_0$ on conspicuous consumption by all players in all Simultaneous Move Rounds (R1-10). Jones (left) vs Veblen (right).
    }
    \label{fig:Scatter_Earnings_ConspCons_r11-15}
\end{figure}

\begin{figure}[H]
    \centering
\begin{subfigure}[t]{0.49\textwidth}
\includegraphics[width=1\linewidth]{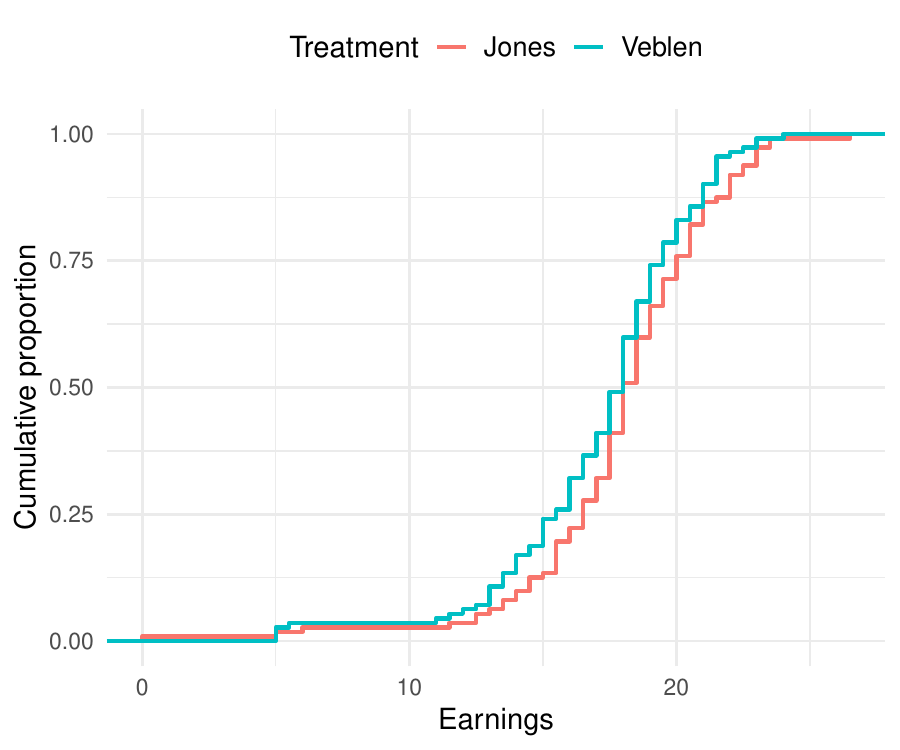}
\end{subfigure}
\begin{subfigure}[t]{0.49\textwidth}
    \includegraphics[width=1\linewidth]{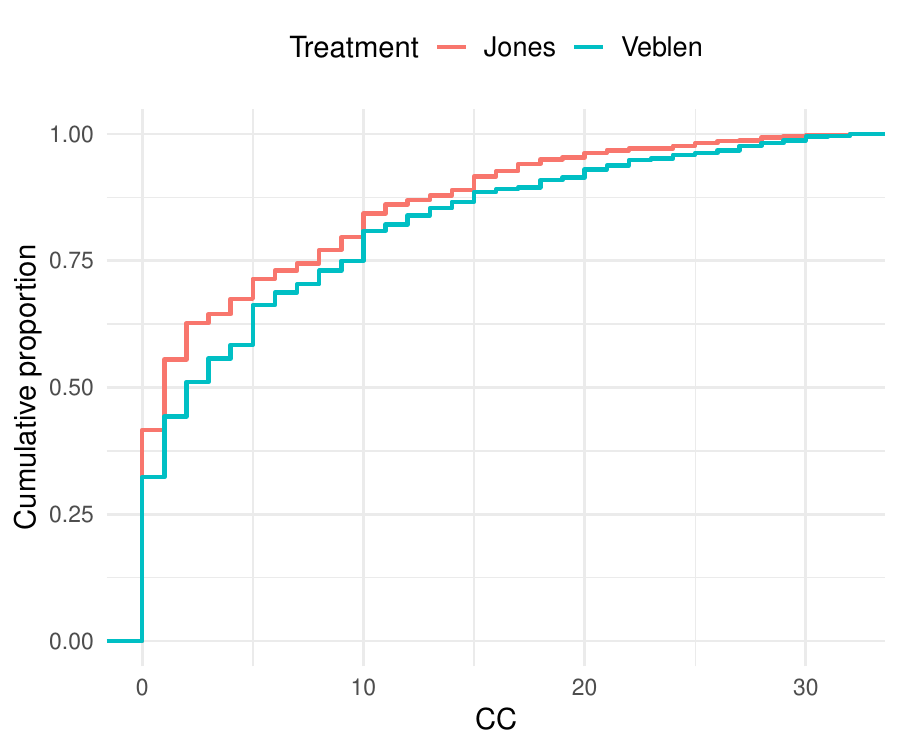}
\end{subfigure}

\caption{Cumulative Distribution Functions (cdf's) for earnings (left) and conspicuous consumption CC (right) in R11-15.}
    \label{fig:cdfs4ConspConsEarningsR11-15}
\end{figure}

Figure 
\ref{fig:Scatter_Earnings_ConspCons_r11-15} shows scatterplots of income and conspicuous consumption for rounds R11-15, 
and 
Figure \ref{fig:cdfs4ConspConsEarningsR11-15} shows the corresponding cumulative distribution functions.
Real effort income is somewhat smaller in R11-15; the median is about 18 tokens versus about 25 tokens in R1-10, perhaps because of fatigue.
The cdf's confirm what was noted in the text: 
a larger fraction of players engage in positive and in serious CC in R11-15, especially in the Veblen treatment.

Tables \ref{tab:Regression_VeblenEffect2} and \ref{tab:RDD_invest3} confirm that the regression results reported in Tables \ref{tab:Regression_VeblenEffect} and \ref{tab:RDD}
are qualitatively similar when the sample of positive CC is further restricted to include only serious CC, i.e., observations where final CC is strictly more than 3 tokens. 

\begin{table}[H]
\begin{center}
\begin{tabular}{l c c c c}
\hline
  & CC  & CC   & Only $CC>3$ & $I_{CC>3}$ \\
  \hline
(Intercept)                  & $0.76$        & $0.97$        & $8.96^{**}$   & $10.53$        \\
                             & $(0.90)$      & $(0.93)$      & $(3.26)$      & $(6.61)$       \\
earnings                     & $0.09^{*}$    & $0.09^{*}$    & $0.18$        & $0.47^{\circ}$ \\
                             & $(0.04)$      & $(0.04)$      & $(0.13)$      & $(0.25)$       \\
Veblen                       & $0.94$        & $0.53$        & $-0.07$       & $4.46$         \\
                             & $(0.76)$      & $(0.82)$      & $(1.62)$      & $(5.11)$       \\
R6-10                        & $-1.02^{***}$ & $-1.43^{***}$ & $-2.57^{*}$   & $-8.28^{***}$  \\
                             & $(0.25)$      & $(0.31)$      & $(1.01)$      & $(1.84)$       \\
Veblen $\times$ R6-10        &               & $0.82^{*}$    & $3.23^{*}$    & $2.47$         \\
                             &               & $(0.38)$      & $(1.30)$      & $(2.60)$       \\
\hline
Num. obs.                    & $2240$        & $2240$        & $467$         & $2240$         \\
F-test: Veblen R6-10 effect = 0 & n/a          & F = 2.449    & F = 0.408       & F = 3.976     \\
p-value                         & n/a          & 0.118        & 0.523           & 0.046         \\
\hline
\multicolumn{5}{l}{\scriptsize{$^{***}p<0.001$; $^{**}p<0.01$; $^{*}p<0.05$; $^{\circ}p<0.1$}}
\end{tabular}
\caption{Same as in Table \ref{tab:Regression_VeblenEffect} but with positive CC replaced by serious CC, i.e., CC $>3$ tokens.}
\label{tab:Regression_VeblenEffect2}
\end{center}
\end{table}

\begin{table}[H]
\begin{center}
\begin{tabular}{l c c c}
\toprule
 & CC & $I_{CC>3}$ & Only CC$>3$\\
\midrule

(Intercept)                   & $2.46^{***}$ & $18.13^{***}$ & $12.45^{***}$ \\
                              & $(0.46)$     & $(3.10)$      & $(1.09)$      \\
R11-15       & $2.04^{**}$  & $17.41^{***}$ & $-0.71$       \\
                              & $(0.65)$     & $(4.34)$      & $(1.19)$      \\
Veblen              & $0.90$       & $5.45$        & $1.24$        \\
                              & $(0.76)$     & $(4.68)$      & $(1.71)$      \\
R11-15$\times$Veblen  & $0.37$       & $3.30$        & $-0.84$       \\
                              & $(0.88)$     & $(6.15)$      & $(1.52)$      \\
\midrule
Num. obs.                     & $3360$       & $3360$          & $1397$          \\
\bottomrule
\multicolumn{4}{l}{\scriptsize{$^{***}p<0.001$; $^{**}p<0.01$; $^{*}p<0.05$; $^{\circ}p<0.1$}}
\end{tabular}
\caption{Same as in Table \ref{tab:RDD} but with CC $>0$ replaced by serious CC, i.e., CC $>3$ tokens.}
\label{tab:RDD_invest3}
\end{center}
\end{table}

\begin{table}[H]
\centering
\begin{tabular}{l|rr|rrrr}
  \toprule
Category & R1 & R10 & Init R11 & Final R11 & Init R15 & Final R15 \\ 
  \midrule
No CC & 0.652 & 0.732 & 0.273 & 0.258 & 0.406 & 0.359 \\ 
Trivial CC ($\leq 3$) & 0.089 & 0.103 & 0.281 & 0.305 & 0.219 & 0.266 \\ 
Serious CC ($>$3) & 0.259 & 0.165 & 0.445 & 0.438 & 0.375 & 0.375 \\ 
   \bottomrule
\end{tabular}
\caption{Conspicuous Consumption Shares in Chosen Rounds. Init Rx refers to initial choices of CC, before Last Move adjustment, in Round x. 
Final Rx refers to $y^* =$ post-adjustment CC .
Rows respectively report the fractions of the 
224 choices in the given round that are zero, 1-3, and more than 3 tokens. 
}
\label{tab:InitVsFinalCC}
\end{table}

Does the jump in CC in the first Last Move period  come mainly from actual last move adjustments?
Table \ref{tab:InitVsFinalCC} indicates otherwise.
It shows that in Round 1, about 9\% 
of players spend a trivial amount on CC, while about  26\% are serious consumers, spending more than 3 tokens. 
By Round 10, the share of serious consumers falls to  16.5\%, but then it jumps to 44.5\% in Round 11 even before Last Move adjustments. 
The share of trivial consumers also jumps, from about 10\% to 28\%.
Last Move adjustments per se have little impact on these shares in Round 11 or even in Round 15. The upshot is that the Last Move treatment draws more active players into the conspicuous consumption game, even before they make final adjustments. 
The share of non-consumers falls immediately to under 30\% and remains well under 40\% even in the last round. 
Equally important, the share of serious consumers 
remains in the vicinity of 40\% both before and after Last Move adjustments.




\begin{table}[H]
\begin{center}
\begin{tabular}{l ccc ccc}
\hline
 & \multicolumn{3}{c}{Active Revisions} & \multicolumn{3}{c}{CC$>3$} \\
\cline{2-4} \cline{5-7}
 & All & Veblen & Jones & All & Veblen & Jones \\
\hline
$w_0$   & $-0.12$     & $-0.33$    & $0.02$     
        & $0.21$    & $0.35$   & $0.26$           \\
        & $(0.19)$   & $(0.25)$   & $(0.42)$     
       & $(0.07)$      & $(0.09)$      & $(0.08)$      \\
$\bar y$ & $1.40^{**}$ & $1.20^{*}$ & $1.49^{**}$  
        & $0.94^{***}$  & $1.06^{***}$  & $1.11^{*}$   \\
        & $(0.42)$    & $(0.55)$   & $(0.44)$ 
        & $(0.20)$      & $(0.12)$      & $(0.45)$      \\
$L$     & $-1.01^{*}$ & $-0.53$    & $-1.88^{***}$
        & $-0.79^{***}$ & $-0.71^{***}$ & $-1.18^{***}$ \\
        & $(0.47)$    & $(0.60)$   & $(0.28)$ 
        & $(0.16)$      & $(0.08)$      & $(0.28)$      \\
\hline
Num. obs.   & $63$   & $39$   & $24$ 
            & $243$  & $135$  & $108$ \\
\hline
\multicolumn{7}{l}{\scriptsize{$^{***}p<0.001$; $^{**}p<0.01$; $^{*}p<0.05$; $^{\circ}p<0.1$}}
\end{tabular}
\caption{Coefficient estimates for equation (\ref{eq:MainRegression}). Left three columns restrict the R11-15 sample to instances where the player changed CC in the Last Move (Active Revisions); right three columns include all R11-15 observations with serious CC ($y^*>3$) and $y^*<m_0$.
}
\label{tab:EnvyPrideRegressionCombined}
\end{center}
\end{table}

The next several tables check the robustness of the key regression results for equation (\ref{eq:MainRegression}) reported in Table \ref{tab:EnvyPrideRegression}.
The first three columns of Table \ref{tab:EnvyPrideRegressionCombined} show that the coefficient estimates 
don't change drastically when we restrict the sample to instances with actual adjustment in CC by the last mover although, of 
course, statistical significance is much lower due to the much smaller number of observations, reflecting behavioral inertia. 
The last 3 columns show that excluding trivial (1-3 token) instances of CC consumption from the main sample still yields coefficient estimates near 1.0 for mean peer consumption $\bar{y}$ and very negative estimates for the dispersion proxy $L$, all significant at $p=0.01$ level, while income effect estimates remain near zero.

Table \ref{table:coefficientsx} uses Block 1 data to obtain an analogue of our main regression results. 
Of course, in rounds 1-10 players do not have the opportunity to best respond contemporaneously, but
they do observe peers' CC distribution in the 
previous period and may choose to best respond to that. 
Accordingly, using data from R2-10 
we regress CC on current income and one period lagged peers' mean $\bar{y}_{-1}$ and $L_{-1}$.
Results are not dissimilar to those reported in Table \ref{tab:EnvyPrideRegression}.
\begin{table}[H]
\begin{center}
\begin{tabular}{l c c c }
\hline
 & All & Veblen & Jones \\ 
\hline
$w_0$                   & $0.10$        & $0.22$        & $-0.09$   \\ 
                                      & $(0.11)$      & $(0.18)$      & $(0.09)$  \\ 
$\bar{y}_{-1}$           & $1.30^{***}$  & $1.22^{***}$  & $1.39^{***}$   \\ 
                                      & $(0.17)$      & $(0.28)$      & $(0.21)$     \\ 
$L_{-1}$                          & $-1.39^{***}$ & $-1.31^{***}$ & $-1.51^{***}$  \\ 
                                      & $(0.15)$      & $(0.20)$      & $(0.22)$    \\ 
\hline
Num. obs.                             & $409$         & $234$         & $175$      \\ 
\hline
\multicolumn{4}{l}{\scriptsize{$^{***}p<0.001$; $^{**}p<0.01$; $^{*}p<0.05$; $^{\cdot}p<0.1$}}
\end{tabular}
\caption{Main regression adapted to Block 1 data. 
Dependent variable is CC when serious ($y_0>3$). 
Explanatory variables include income ($w_0$) and lagged peers' CC mean $\bar{y}_{-1}$ and lagged upward comparison $L_{-1}$.
Column 1 is for all R2-10 data, while Columns 2 and 3 are for the Veblen and Jones subsamples respectively.
}
\label{table:coefficientsx}
\end{center}
\end{table}

The next two Tables look at our most conservative specification for estimating equation (\ref{eq:MainRegression}):
the data include only active changes and the dispersion measures are conventional
and ignore the focal player's choices.
The dispersion variable coefficients remain negative but due to small sample sizes are at best marginally significant.

\begin{table}[H]
\begin{center}
\begin{tabular}{l c c c}
\hline
 & All & Veblen & Jones \\
\hline
$w_0$                    & $-0.26$         & $-0.38$     & $-0.03$  \\
                                      & $(0.18)$        & $(0.25)$    & $(0.33)$ \\
$\bar{y}$                    & $0.76^{**}$     & $0.79^{**}$ & $1.54$   \\
                                      & $(0.23)$        & $(0.27)$    & $(1.35)$ \\
$y_1 - y_3$                 & $-0.34^{\circ}$ & $-0.15$     & $-0.84$  \\
                                      & $(0.18)$        & $(0.25)$    & $(0.64)$ \\
\hline
Num. obs.                             & $63$            & $39$        & $24$     \\
\hline
\multicolumn{4}{l}{\scriptsize{$^{***}p<0.001$; $^{**}p<0.01$; $^{*}p<0.05$; $^{\circ}p<0.1$}}
\end{tabular}
\caption{OLS coefficient estimates for equation (6). Data are the same as for first three columns in Table \ref{tab:EnvyPrideRegressionCombined}, but dispersion proxy $L$ is replaced by the range width of peers' CC, $y_1 - y_3$. }
\label{table:active-spread}
\end{center}
\end{table}

\begin{table}[H]
\begin{center}
\begin{tabular}{l c c c}
\hline
 & All & Veblen & Jones \\
\hline
$w_0$                      & $-0.26$         & $-0.38$     & $-0.03$  \\
                                      & $(0.18)$        & $(0.25)$    & $(0.33)$ \\
$\bar{y}$                             & $0.75^{**}$     & $0.80^{**}$ & $1.16$   \\
                                      & $(0.22)$        & $(0.26)$    & $(1.13)$ \\
$Std[y_{-0}]$                           & $-0.65^{\circ}$ & $-0.30$     & $-1.25$  \\
                                      & $(0.33)$        & $(0.46)$    & $(1.00)$ \\
\hline
Num. obs.                             & $63$            & $39$        & $24$     \\
\hline
\multicolumn{4}{l}{\scriptsize{$^{***}p<0.001$; $^{**}p<0.01$; $^{*}p<0.05$; $^{\circ}p<0.1$}}
\end{tabular}
\caption{OLS coefficient estimates for equation (6). Data are the same as for first three columns in Table \ref{tab:EnvyPrideRegressionCombined}, but dispersion proxy $L$ is replaced by standard deviation of peers' CC.}
\label{table:active-std}
\end{center}
\end{table}


\begin{figure}[H]
    \centering    \includegraphics[width=0.5\textwidth]{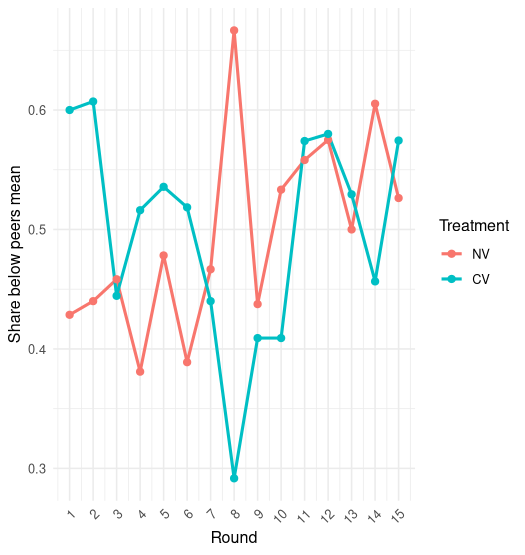}
\caption{Fraction of positive conspicuous consumers with CC  below peers' mean.}
    \label{fig:BelowAvg}
\end{figure}
The PT-MEP model predicts that consumers will either choose $y^* = 0$ or else $y^* \geq \bar{y}$, and never $y^* \in (0, \bar{y}).$
Figure \ref{fig:BelowAvg} shows to the contrary that around half the positive CC choices are in the forbidden zone, and if anything that fraction trends upward in later periods.

\begin{figure}[H]
    \centering
    \includegraphics[width=0.5\textwidth]{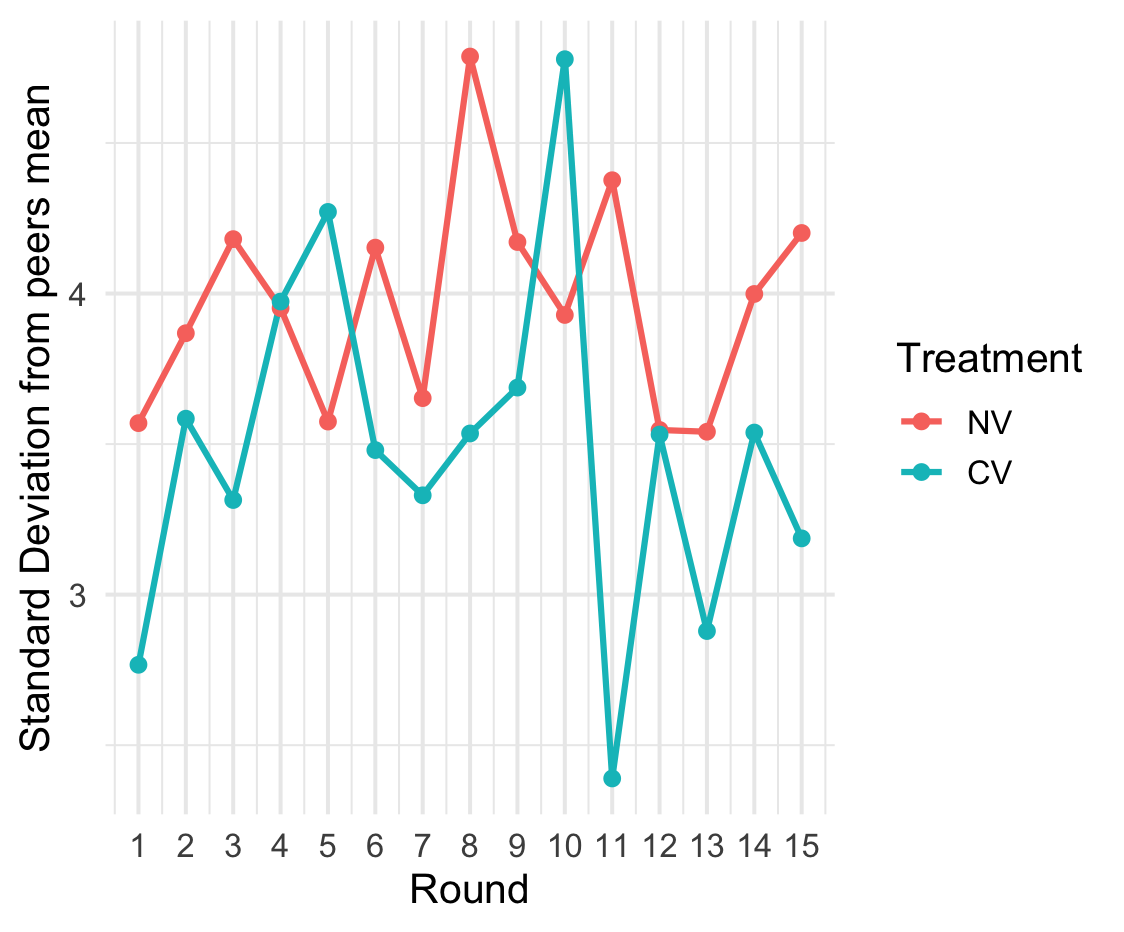}
    \caption{Standard Deviation of positive CC.}
    \label{fig:StdDevTrends}
\end{figure}
The MEP model with parameters satisfying $c_e > 1 > c_p >0$ predicts that everyone aims for average CC, and so we should see the dispersion around $\bar{y}$ decrease over time. 
Figure \ref{fig:StdDevTrends} shows that, to the contrary, standard deviation among positive conspicuous consumers has no tendency to decline from one round to the next; if anything, there is a noisy upward trend for both Jones and Veblen treatments.

\newpage
\section{Mathematical details}
\subsection{Rank convention and ties}\label{ssec:AppRDManal}

Recall that peers are indexed so that
\[
y_1 \geq y_2 \geq \cdots \geq y_N \geq 0,
\]
where $y_0$ denotes the focal player's choice. Our convention assigns the
focal player rank
\[
r(y_0,y_{-0})
=
1+\sum_{i=1}^{N}\mathbf{1}\{y_i\geq y_0\}.
\tag{B.1}
\]
Thus every peer whose conspicuous consumption is weakly greater than that
of the focal player is ranked ahead of her. In particular, a tie with a peer
does not allow the focal player to share that peer's higher rank. Since
$y_i\geq 0$ for all $i$, equation (B.1) also implies
$r(0,y_{-0})=N+1$.

For $y_0>0$, this definition is equivalent to the interval convention used
in the main text. Setting $y_{N+1}=0$, rank $r$ is obtained whenever
\[
y_0\in (y_r,y_{r-1}],
\tag{B.2}
\]
with the natural convention that the upper endpoint for rank $1$ is
$+\infty$. The interval is open at its lower endpoint because choosing
$y_0=y_r$ leaves peer $r$ weakly ahead of the focal player. Hence, to move
above that peer, the focal player must choose strictly more than $y_r$.

Importantly, the ordering of peer choices is weak, so peers themselves may
be tied. If
\[
y_{r-1}=y_r,
\]
then
\[
(y_r,y_{r-1}]=\varnothing,
\]
and rank $r$ is not attainable. Thus ties among peers can cause the focal
player to skip one or more ranks. More generally, suppose that $m$ peers
choose the same level $x$. Moving from $y_0\leq x$ to $y_0>x$ overtakes
all $m$ of these peers simultaneously and therefore improves the focal
player's rank by $m$ positions.

For example, if the three peers choose
\[
(y_1,y_2,y_3)=(20,15,15),
\]
then the focal player's possible ranks satisfy
\[
r(y_0,y_{-0})=
\begin{cases}
1, & y_0>20,\\
2, & 15<y_0\leq20,\\
4, & 0\leq y_0\leq15.
\end{cases}
\]
Rank $3$ is unattainable because its nominal interval is
$(15,15]=\varnothing$. In particular, choosing $y_0=15$ gives rank $4$,
whereas choosing any $y_0>15$ jumps directly to rank $2$.

\paragraph{Rank-dependent utility.}
In the Rank-dependent model,
\[
\phi(y_0,y_{-0})=C(r(y_0,y_{-0})),
\]
where we maintain
\[
C(1)\geq C(2)\geq\cdots\geq C(N+1).
\tag{B.3}
\]
Thus social subutility depends only on the focal player's ordinal position
and not on the cardinal distance between her CC and the choices of her
peers. For the sharp version of the Rank-dependent prediction, we impose
strict monotonicity,
\[
C(1)>C(2)>\cdots>C(N+1).
\tag{B.4}
\]
Strict monotonicity means that an improvement in attainable rank produces
a strictly higher social payoff. It does not imply that the player always
chooses the highest rank, since achieving a higher rank requires greater
expenditure.

Conditional on attaining a given rank, additional expenditure has no
status benefit. Indeed, on every nonempty constant-rank interval,
\[
\frac{\partial \phi(y_0,y_{-0})}{\partial y_0}=0,
\]
whereas the marginal cost of CC is one. Since
\[
U(y_0,y_{-0})
=
w_0-y_0+C(r(y_0,y_{-0})),
\]
utility therefore decreases one-for-one with $y_0$ as long as rank remains
unchanged. Consequently, conditional on targeting a particular attainable
rank, the player chooses the least costly CC level that achieves that rank.

To state this formally for the experiment, where CC is measured in
indivisible units, define the set of attainable ranks
\[
\mathcal{R}(y_{-0})
=
\{1\}
\cup
\bigl\{
r\in\{2,\ldots,N\}:y_{r-1}>y_r
\bigr\}
\cup
\{N+1\}.
\tag{B.5}
\]
For every positive attainable rank $r\leq N$, the minimum expenditure
required to achieve that rank is
\[
q_r=y_r+1,
\tag{B.6}
\]
while
\[
q_{N+1}=0.
\tag{B.7}
\]
For rank $1$, equation (B.6) is interpreted as
$q_1=y_1+1$. Subject to affordability, the Rank-dependent best-response
problem can therefore be written as the finite problem
\[
r^*
\in
\arg\max_{\substack{
r\in\mathcal{R}(y_{-0})\\
q_r\leq w_0}}
\left\{
C(r)-q_r
\right\},
\tag{B.8}
\]
with
\[
y_0^*=q_{r^*}.
\tag{B.9}
\]
The term $w_0$ can be omitted from (B.8) because it is common to all
feasible choices.

Equations (B.5)--(B.9) also clarify the role of peer ties. If several peers
are tied at $y_r$, spending one additional unit,
\[
y_0=y_r+1,
\]
may overtake several peers simultaneously. The focal player then jumps to
the corresponding attainable rank rather than occupying any of the
intermediate, unattainable ranks. Thus the statement that a
Rank-dependent player chooses ``one unit above the next lower peer'' should
be understood as a statement about the minimum expenditure required to
cross an \emph{attainable rank threshold}, rather than as requiring every
integer rank $1,\ldots,N+1$ to be attainable.

Notice also that our tie convention does not imply that an optimal choice
can never numerically equal another peer's choice. For example, if peers
choose $(20,15,14)$, the least costly choice attaining rank $3$ is
$y_0=15$: the focal player ties the peer choosing $15$, who remains ranked
ahead of her, while she strictly exceeds the peer choosing $14$. What the
convention rules out is obtaining a higher rank \emph{by tying} the player
who must be overtaken.

With a continuous choice space, there need not be a smallest value strictly
above $y_r$ because the lower endpoint of $(y_r,y_{r-1}]$ is open. In that
case the Rank-dependent model predicts a choice arbitrarily close to
$y_r$ from above. In our experiment, indivisibility resolves this issue:
the smallest possible increment is one token, yielding the sharp prediction
$y_0^*=y_r+1$ whenever a positive attainable rank $r$ is targeted.


\subsection{PEP Analysis}\label{ssec:AppPEPanal}
\PEPcases*

\vspace{-.1in}
\noindent
\textbf{Proof.} Recall that peers are indexed so that $y_1 \geq y_2 \geq ... \geq y_N\geq 0$, and that rank $r \leq N+1$ is the positive integer such that $y_0 \in (y_r, y_{r-1}]$. 
Differentiating PEP subutility 
$\phi(y_0, \cdot) = \frac{c_e}{N} \sum_{i=1}^{r-1} [y_0 -y_i ] + \frac{c_p}{N} \sum_{i=r-1}^N [y_0 -y_i]$
with respect to CC choice $y_0$,
we have 
\begin{equation}\label{eq:phi1}
 \phi_1 = c_e \frac{r-1}{N} + c_p \frac{N-r+1}{N},   
\end{equation}
a weighted average of $c_e$ and $c_p$.
Thus $\phi_1 >1$ (resp $<1$) everywhere when $c_e, c_p >1$ (resp $<1$) so the best response is the corner solution
$y^* = w_0$ (resp $y^* = 0$),
establishing parts 1 and 2.

\vspace{-.05in}

For parts 3 and 4, note that by equation (\ref{eq:phi1}) the incentive $\frac{dU}{dy_0}= -1 + \phi_1$ to increase $y_0$ is constant within each interval of constant rank
$J_r \equiv (y_r, y_{r-1}]$, and jumps by $\frac{c_e - c_p}{N}$ when $y_0$ moves from interval $J_r$ to $J_{r+1}.$ 
Treating 
 $\alpha = \frac{r-1}{N}$ as a continuous variable,
the FOC can be written as 
$1 = \phi_1 = \alpha c_e + (1-\alpha) c_p$, with solution $\alpha^* = \frac{1-c_p}{c_e - c_p} \in (0,1).$
In case 3, the solution $\alpha^* = \frac{1-c_p}{c_e - c_p} \in (0,1)$ to the FOC points to a relative minimum, not to the best response. 
Equation (\ref{eq:phi1}) shows that
moving $y_0$ up from interval $J_r$ where $r \approx N\alpha^* +1$ and $\phi_1 \approx 1$ to interval $J_{r-1}$ and beyond will further increase 
$\phi_1$, and strengthen the incentive to increase $y_0$ further. 
Moving it below $y_{r}$ will push the $\phi_1$ further below 1, and strengthen the incentive to further decrease $y_0$. 
Thus in case 3, we  have a local max at the corner $y_0  = w_o$ and another local max at the other corner $y_0=0$.
Which corner is the global max depends on the distribution of peers' choices and on the exact values of $c_e , c_p$.

The most interesting (and perhaps the most realistic) case is 4, where $c_e > 1> c_p>0.$
Here the best response is interior, and characterized by $\alpha^* = \frac{1-c_p}{c_e - c_p} \in (0,1)$.
The interpretation is that the player (a) targets a particular rank $r \approx \alpha^* N +1$  that best balances pride and envy, and
(b) has almost flat preferences within that rank.
To spell this out, at the targeted $r$ we have $\phi_1<1$ for $y_0 \geq y_{r-1}$
and $\phi_1>1$ for $y_0 \leq y_{r}.$
Equation (\ref{eq:phi1}) tells us that the values of $\phi_1$ at the endpoints of $J_r$ differ by $\frac{c_e - c_p}{N}$ so they differ from 1 by even less.
Equation (\ref{eq:phi1}) also tells us that incentive $\phi_1-1$ to move towards the top of the targeted interval $J_r$ is constant.
The incentive can be arbitrarily close to zero and is bounded above by 
$\frac{c_e - c_p}{N}>0$ and bounded below by $\frac{c_p - c_e}{N}<0$.
The intuition is that
the marginal benefit of increasing $y_0$ decreases when there are fewer peers to envy. \qed

\subsection{PT-PEP analysis}\label{ssec:AppPTPEPanal}
\PTPEPpred*
\noindent
\textbf{Proof sketch.}
Recall that the marginal benefit function for the PT-PEP model is 
\begin{equation}\label{eq:ptpepmb}
    \phi_1 = \frac{1}{N}\sum_{i=1}^N V'(y_0 - y_i),
\end{equation}
where the value function $V$ is S-shaped and kinked: $V'(x)>0$ with $xV''(x)<0$ for $x\neq 0$, while $V'(0-)=c_e>c_p=V'(0+)$ 
as in Figure \ref{fig:PTvalfnGrad}.
Given the vector $y_{-0}$ of peers' CC, 
equation (\ref{eq:ptpepmb}) tells us that $\phi_1(y_0)$ is positive, and is continuous everywhere except at $y_0 = y_i$ for each peer  $i$, where it has discontinuous down jump of
$\frac{c_p -c_e}{N}<0$. 
The rest of this sketch replaces $\phi_1$ by an arbitrarily close smooth approximation. 

The smoothed function $\phi_1(y_0)$ attains a maximum value $M>0$ on the compact set $[0, w_0]$.
If $M<1$ then the marginal benefit is always less than the marginal cost, and the best response is $y^* =0$.
Likewise, if $\phi_1(y_0)>1$ everywhere
then $y^* = w_0$.
Otherwise, by the intermediate value theorem, there are solutions to the FOC $\phi_1=1$. 
Some of these may be upcrossings, where $\phi_{11}>0$; such solutions are local minima. Downcrossings, where $\phi_{11}<0$, are local maxima. 
A global best response $y^*$ will either be a corner (0 or $w_0$) or else is a local maximum, where $\phi_1(y^*)=1$ 
and $\phi_{11}(y^*)<0.$

We now show that properties (a) and (b) of the Prediction hold at such interior solutions. If we shift each peer's CC $y_i$ to $y_i +z$ then the arguments $(y_0 - y_i)$ in equation (\ref{eq:ptpepmb}) are unaffected when we shift own CC $y_0$ in parallel to $y_0 +z$. Hence, as long as it remains interior, $y^* +z$ is a best response to $y_{-0}+z$ whenever $y^*$ is a best response to $y_{-0}$, and property (a) follows.

To establish property (b), let $F(y_0,y_{-0}) = -1+\phi_1$ denote the net incentive, so the FOC
 $F(y^*,y_{-0})=0$ and the downcrossing expression $F_{y_0}(y^*,y_{-0})=\frac1N\sum_{i=1}^N V''(y^*-y_i)<0$ hold at an interior 
 best response $y^*$.
Implicitly differentiating the FOC, we obtain
\begin{equation}\label{eq:Implicit}
 \frac{\partial y^*}{\partial y_j}
= -\frac{F_{y_j}}{F_{y_0}}
= \frac{V''(y^*-y_j)}{\sum_{i=1}^N V''(y^*-y_i)}.   
\end{equation}
The denominator in (\ref{eq:Implicit}) is negative since we are at a downcrossing, and the S-shape property ($xV''(x)<0$ for $x\neq 0$) tells us that the numerator has the opposite sign of $x=y^*-y_j$. Therefore
\begin{equation}\label{eq:pwmps}
\frac{\partial y^*}{\partial y_j} \lessgtr 0 \mbox{ as }  y^* \lessgtr y_j.
\end{equation}
That is, the best response $y^*$ is decreasing in the CC of leaders (peers $j$ s.t. $y^*<y_j$) and increasing in the CC of laggards ( $y^*>y_j$). 
Thus a pairwise mean preserving spread that increases a leader's CC offset by a decrease in a laggard's CC will doubly decrease $y^*$, so E.2 follows.
\qed

\subsection{Income Effects in Multiplicative models}\label{ssec:AppMultM}

We begin with a simple example drawn from \cite{bramoulle2024status}, 
where the CC subutility can be written $\phi(y_0, y_{-0}) = y_0-\frac{\alpha}{1+\alpha}\bar{y}$.
To make the point without needless complications, take $\alpha = 0.5$ and consider only the subsample of data where $\phi>0,$
i.e., where $y_0 > \bar{y}/3$.
By convention, the price of CC is $p=1,$ so
the budget constraint can be written $m_0=w_0-y_0$, 
where $w_0$ is earned income and $y_0$ is the expenditure on CC.
BG24's multiplicative objective function then is
$U(y_0,\mathbf{y}_{-0})=(w_0-y_0)^{\sigma}(y_0-\bar{y}/3)^{1-\sigma}$ with $y_0 \in (\bar{y}/3, w_0]$ and 
Cobb-Douglas parameter
$\sigma \in (0,1)$. 
Since  $\frac{\partial U}{\partial y_0}=
\frac{-\sigma U}{w_0-y_0}+\frac{(1-\sigma)U}{y_0-\bar{y}}$, the
FOC $\frac{\partial U}{\partial y_0}=0$  implies $\sigma( y_0-\bar{y}/3) = (1-\sigma)(w_0-y_0),$ which
yields interior best response
$y^*=(1-\sigma)w_0 +\sigma  \bar{y}/3$. 
Thus the income effect in this example is $\frac{dy^*}{dw_0} = 1-\sigma >0$.

Using a more general CC subutility function $\phi(y_0,y_{-0}) \geq 0$,
the multiplicative form
$M(m_0,y_0, y_{-0})=m_0^\sigma \phi(y_0,y_{-0})^{1-\sigma}$ has FOC 
$0=\frac{\partial M}{\partial y_0}=
\frac{-\sigma M}{w_0-y_0}+\frac{(1-\sigma)M\phi_1}{\phi}$. 
Thus interior best responses $y^*=y_0$ satisfy 
\begin{equation}\label{eq:MultFOC}
\sigma \phi= (w_0 - y^*) (1-\sigma)\phi_1 .
\end{equation}
Implicitly differentiating (\ref{eq:MultFOC}) wrt $w_0$,
we get income effect $\frac{dy^*}{dw_0} = \frac{(1-\sigma)\phi_1}{\phi_1 - (w_0-y^*)(1-\sigma)\phi_{11}}$.
The effect is strictly positive given our maintained assumption is that $\phi$ is increasing  in own CC since the second order condition ensures that the denominator is positive.


By contrast, as we have seen, the quasi-linear form
$Q(m_0,y_0, y_{-0})=m_0 + \phi(y_0,y_{-0})$ for the same given CC subutility $\phi$ has FOC 
$\phi_1 = 1$.  
Thus in quasi-linear models there is no income effect at an interior best response.

\subsection{Binomial test details}\label{ssec:bintest}
The test is for an sample of $n$ observations where active adjustment changes ordinal waste $W$, of which $a$ observations are in the wrong direction ($W_F > W_I$) and $b=n-a$ are in the intended direction ($W_F < W_I$, waste reduction).
Thus the observed error rate is $\hat{q}=a/n$.

The critical error rate $q^* \in [0.1]$ is defined as the Bernoulli parameter such that at least $a$ failures will be observed with probability 0.05 in $n$ independent Bernoulli trials.  
That is, under the usual conventions, we can confidently reject the hypothesis that adjustments are intended to reduce waste but are subject to directional errors of frequency $q^*$ or less. 
Informally, the data would seem to be compatible with an intention to reduce ordinal waste if $q^*$ is not too large, say 0.10.
Of course, $q^*>0.5$ would indicate that we can be confident that players are perversely trying to \textit{increase} waste.

An analytic expression is 
$q^* = F^{-1}_{\mathrm{Beta}(a,b+1)}(0.05)$. 
More explicitly, $q^*$ is the solution in $(0,1)$ to 
\begin{equation}
\sum_{k=a}^{n} \binom{n}{k} q^k (1-q)^{n-k} = 0.05.   
\end{equation}

For example, consider the first line in Table \ref{tab:OrdinalTheories} where $n=20, a=11, b=9.$
Evaluating numerically, we find $\sum_{k=11}^{20} \binom{20}{k} (0.347)^k (0.653)^{20-k} = 0.050069$,
indicating that the given value $q^* = 0.347$ is accurate up to the three given digits.

\clearpage

\section{Experimental Instructions (Online)}
\label{sec:ExperimentalInstructions}

\subsection{Jones (initial instructions, read before round 1)}
\input{instructions/NV_initial}

\subsection{Jones (additional instruction, read after round 10)}
\input{instructions/NV_additional}

\subsection{Veblen (initial instructions, read before round 1)}
\input{instructions/CV_initial}

\subsection{Veblen (additional instruction, read after round 10)}
\input{instructions/CV_additional}

\newpage
\bibliography{refs}

\end{document}

%% file: preambleR.tex
\usepackage[english]{babel}
\usepackage[T1]{fontenc}

\usepackage[margin=1.3in]{geometry}

\usepackage{setspace}
\usepackage{amsmath}
\usepackage{amssymb}

\usepackage{newtxtext,newtxmath}
\usepackage{adjustbox}
\usepackage{graphicx}
\usepackage{rotating}

\usepackage[colorlinks=true, allcolors=blue]{hyperref}

\usepackage{amsthm}
\usepackage{lscape}
\usepackage{thmtools}
\usepackage{thm-restate}

\usepackage{color,soul}
\usepackage{xcolor}

\usepackage{tikz}

\usepackage{booktabs}
\usepackage{booktabs, caption}
\usepackage{tabularx}
\usepackage{array}
\newcolumntype{P}[1]{>{\centering\arraybackslash}p{#1}}
\newcolumntype{M}[1]{>{\centering\arraybackslash}m{#1}}
\usepackage[flushleft]{threeparttable}
\usepackage{multirow}
\usepackage{siunitx}
\usepackage{changepage}
\usepackage{float}
\usepackage{placeins}
\usepackage{rotating} 
\usepackage{natbib}
\setcitestyle{authoryear,open={(},close={)}}

\usepackage[toc,page]{appendix}

\def\sym#1{\ifmmode^{#1}\else\(^{#1}\)\fi}

\newtheorem{hyp}{Hypothesis}
\newtheorem{prediction}{Prediction}

\usepackage{soul}

\usepackage{pdfpages}

 \usepackage{mathtools}
 
\newcounter{marginparcounter}

\usepackage{graphicx} 
\usepackage{subcaption} 
\usepackage{rotating} 

\usepackage[utf8]{inputenc}
\usepackage{eurosym}

\usepackage{booktabs}
\usepackage{quotes}
\usepackage[table]{xcolor}
\usepackage{amsmath,amsthm}
\usepackage{amssymb}
\usepackage{bbm,multirow}
\usepackage{mathtools}

\usepackage{textcomp}
\usepackage{caption,subcaption}

\usepackage{tikz,pgfplots}

\newcommand{\R}{\mathbb{R}}

\newcommand{\red}{\textcolor{red}}

\usepackage{xpatch}
\makeatletter
\AtBeginDocument{\xpatchcmd{\@thm}{\thm@headpunct{.}}{\thm@headpunct{}}{}{}}
\makeatother

\usepackage{pifont}

\usepackage[normalem]{ulem} 

%% file: instructions/NV_initial.tex
Welcome to the \textbf{ESSEXLab!} You have earned \textbf{£5 for showing up on time}. Please abstain from using any personal electronic devices (including phones, tablets, and headphones) as well as communicating with other players while the experiment is in progress.

\bigskip

\begin{itemize}
    \item You will be playing 10 rounds following the rules explained below.
    \item You can earn up to £30 depending on the round chosen for payment.
    \item Before the experiment begins you will choose a user name that will be your \textbf{virtual identity} throughout the experiment.
\end{itemize}

\subsection*{Structure of a Round}

You will be assigned to a group of four participants, which will remain the same for the entire experiment. Each round, you and the other members of your group will collect decorative objects called decals, and will compare collections. Each round consists of three stages, which are detailed below.

\subsection*{Stage 1}

\begin{itemize}
    \item You will see a set of sliders on your screen.
    \item You can adjust each slider to any position between 0 and 100 by clicking and dragging with your mouse.
    \item You have one minute to adjust as many sliders as possible to the target position of 50.
    \item You will earn one token for each correctly adjusted slider.
\end{itemize}

\subsection*{Stage 2}

\begin{itemize}
    \item You can purchase digital decals at a cost of one token per decal:
    \begin{center}
        \includegraphics[width=0.15\textwidth]{pictures/decal.png}
    \end{center}
    \item Decals have no monetary value and will not affect your final payment.
    \item For example, if you earn 21 tokens from Stage 1 and choose to purchase 10 decals, you will be paid 11 tokens if that round is selected for payment.
\end{itemize}

\subsection*{Stage 3}

\begin{itemize}
    \item After all participants have made their purchasing decisions, you will proceed to the outcome stage.
    \item You will see a table displaying your decal purchase decisions alongside those of three other participants.
\end{itemize}

\begin{center}
    \includegraphics[width=0.8\textwidth]{pictures/NoVisibility.png}
\end{center}

\begin{itemize}
    \item You will then move on to the next round.
\end{itemize}

\subsection*{Payment}

\begin{itemize}
    \item Once the experiment concludes, one round will be randomly selected for payment.
    \item Each round has an equal chance of being chosen, so please take them seriously, as a significant amount of money is at stake.
    \item At the end of the experiment, tokens will be converted to pounds at the following exchange rate:
\end{itemize}

\begin{center}
    \textbf{1 token = 50 pence}
\end{center}

%% file: instructions/NV_additional.tex
Stages 1, 2 and 3, and the payment procedure, are as before, but from now on, each period will include a new stage as follows.

\subsection*{New Stage}

\begin{itemize}
    \item After seeing the outcome of Stage 2, you have the opportunity to revise the number of decals you purchased in Stage 2.
    \item You can increase, decrease or keep the same number of decals.
    \item After you and the other 3 participants made their revisions, the computer will randomly choose to implement the revision of one of the 4 participants.
    \item This means, in every round, there is a 25\% chance that the final outcome table will include your revised choice, but not those of the other 3 participants; their choices would be unchanged.
    \item You will then move to the next round.
\end{itemize}

%% file: instructions/CV_initial.tex
Welcome to the \textbf{ESSEXLab!} You have earned \textbf{£5 for showing up on time}. Please abstain from using any personal electronic devices (including phones, tablets, and headphones) as well as communicating with other players while the experiment is in progress.

\bigskip

\begin{itemize}
    \item You will be playing 10 rounds following the rules explained below.
    \item You can earn up to £30 depending on the round chosen for payment.
    \item Before the experiment begins you will choose a user name that will be your \textbf{virtual identity} throughout the experiment.
\end{itemize}

\subsection*{Structure of a Round}

You will be assigned to a group of four participants, which will remain the same for the entire experiment. Each round, you and the other members of your group will collect decorative objects called decals, and will compare collections. Each round consists of three stages, which are detailed below.

\subsection*{Stage 1}

\begin{itemize}
    \item You will see a set of sliders on your screen.
    \item You can adjust each slider to any position between 0 and 100 by clicking and dragging with your mouse.
    \item You have one minute to adjust as many sliders as possible to the target position of 50.
    \item You will earn one token for each correctly adjusted slider.
\end{itemize}

\subsection*{Stage 2}

\begin{itemize}
    \item You can purchase digital decals at a cost of one token per decal:
    \begin{center}
        \includegraphics[width=0.15\textwidth]{pictures/decal.png}
    \end{center}
    \item Decals have no monetary value and will not affect your final payment.
    \item For example, if you earn 21 tokens from Stage 1 and choose to purchase 10 decals, you will be paid 11 tokens if that round is selected for payment.
\end{itemize}

\subsection*{Stage 3}

\begin{itemize}
    \item After all participants have made their purchasing decisions, you will proceed to the outcome stage.
    \item You will see a table displaying your decal purchase decisions alongside those of your group members.
\end{itemize}

\begin{center}
    \includegraphics[width=0.8\textwidth]{pictures/CardinalVisibility.png}
\end{center}

\begin{itemize}
    \item Other participants will also see your consumption decision.
    \item You will then move on to the next round.
\end{itemize}

\subsection*{Payment}

\begin{itemize}
    \item Once the experiment concludes, one round will be randomly selected for payment.
    \item Each round has an equal chance of being chosen, so please take them seriously, as a significant amount of money is at stake.
    \item At the end of the experiment, tokens will be converted to pounds at the following exchange rate:
\end{itemize}

\begin{center}
    \textbf{1 token = 50 pence}
\end{center}

%% file: instructions/CV_additional.tex
Stages 1, 2 and 3, and the payment procedure, are as before, but from now on, each period will include a new final stage as follows.

\subsection*{New Stage}

\begin{itemize}
    \item After seeing the outcome of Stage 2, you have the opportunity to revise the number of decals you purchased in Stage 2.
    \item You can increase, decrease or keep the same number of decals.
    \item After you and the other 3 members of your group have made their revisions, the computer will randomly choose to implement the revision of one of the 4 group members.
    \item This means, in every round, there is a 25\% chance that the final outcome table will include your revised choice, but not those of the other 3 members of your group; their choices would be unchanged.
    \item You will then move to the next round.
\end{itemize}

%% file: refs.bib
@article{chen2016otree,
  title={oTree—An open-source platform for laboratory, online, and field experiments},
  author={Chen, Daniel L and Schonger, Martin and Wickens, Chris},
  journal={Journal of Behavioral and Experimental Finance},
  volume={9},
  pages={88--97},
  year={2016},
  publisher={Elsevier}
}

@article{abel1990asset,
  title={Asset prices under habit formation and catching up with the Joneses},
  author={Abel, Andrew B},
  journal={American Economic Review, Papers and Proceedings},
  volume={80},
  number={2},
  pages={38--42},
  year={1990}
}

@article{barberis2013thirty,
  title={Thirty years of prospect theory in economics: A review and assessment},
  author={Barberis, Nicholas C},
  journal={Journal of Economic Perspectives},
  volume={27},
  number={1},
  pages={173--96},
  year={2013}
}

@incollection{bramoulle2016games,
  title={Games played on networks},
  author={Bramoull{\'e}, Yann and Kranton, Rachel},
  booktitle={Oxford Handbook of the Economics of Networks},
    year={2016},
  publisher={Oxford University Press}
}

@article{BrownBulteZhang2011,
  author  = {Brown, Philip H. and Bulte, Erwin and Zhang, Xiaobo},
  title   = {Positional Spending and Status Seeking in Rural China},
  journal = {Journal of Development Economics},
  year    = {2011},
  volume  = {96},
  number  = {1},
  pages   = {139--149},
  doi     = {10.1016/j.jdeveco.2010.05.007}
}

@article{charles2009conspicuous,
  title={Conspicuous consumption and race},
  author={Charles, Kerwin Kofi and Hurst, Erik and Roussanov, Nikolai},
  journal={Quarterly Journal of Economics},
  volume={124},
  number={2},
  pages={425--467},
  year={2009},
  publisher={MIT Press}
}

@article{de2020consumption,
  title={Consumption network effects},
  author={De Giorgi, Giacomo and Frederiksen, Anders and Pistaferri, Luigi},
  journal={The Review of Economic Studies},
  volume={87},
  number={1},
  pages={130--163},
  year={2020},
  publisher={Oxford University Press}
}

@book{frank1985choosing,
  title={Choosing the right pond: Human behavior and the quest for status.},
  author={Frank, Robert H},
  year={1985},
  publisher={Oxford University Press}
}

@article{immorlica2017social,
  title={Social status in networks},
  author={Immorlica, Nicole and Kranton, Rachel and Manea, Mihai and Stoddard, Greg},
  journal={American Economic Journal: Microeconomics},
  volume={9},
  number={1},
  pages={1--30},
  year={2017}
}

@book{veblen1899thetheory,
  title={The Theory of the Leisure Class: An Economic Study in the Evolution of Institutions},
  author={Veblen, Thorstein B.},
  year={1899},
  publisher={New York: Macmillan}
}

@article{bertrand2016trickle,
  title={Trickle-down consumption},
  author={Bertrand, Marianne and Morse, Adair},
  journal={Review of Economics and Statistics},
  volume={98},
  number={5},
  pages={863--879},
  year={2016},
  publisher={The MIT Press}
}

@article{campbell1999force,
  title={By force of habit: A consumption-based explanation of aggregate stock market behavior},
  author={Campbell, John Y. and Cochrane, John H.},
  journal={Journal of Political Economy},
  volume={107},
  number={2},
  pages={205--251},
  year={1999}
}

@article{drechsel2014consumption,
  title={Consumption--savings decisions under upward-looking comparisons},
  author={Drechsel-Grau, Moritz and Schmid, Kai D},
  journal={Journal of Economic Behavior \& Organization},
  volume={106},
  pages={254--268},
  year={2014},
  publisher={Elsevier}
}

@book{duesenberry1949income,
  title={Income, Saving and the Theory of Consumer Behavior},
  author={Duesenberry, James S.},
  year={1949},
  publisher={Harvard University Press}
}

@article{frank1985demand,
  title={The demand for unobservable and other nonpositional goods},
  author={Frank, Robert H},
  journal={American Economic Review},
  volume={75},
  number={1},
  pages={101--116},
  year={1985},
  publisher={JSTOR}
}

@article{frank2014expenditure,
year = {2014},
volume = {1},
journal = {Review of Behavioral Economics},
title = {Expenditure Cascades},
number = {1–2},
pages = {55-73},
author = {Robert H. Frank and Adam Seth Levine and Oege Dijk}
}

@article{friedman2008conspicuous,
  title={Conspicuous consumption dynamics},
  author={Friedman, Daniel and Ostrov, Daniel N},
  journal={Games and Economic Behavior},
  volume={64},
  number={1},
  pages={121--145},
  year={2008},
  publisher={Elsevier}
}

@article{ghiglino2010keeping,
  title={Keeping up with the neighbors: social interaction in a market economy},
  author={Ghiglino, Christian and Goyal, Sanjeev},
  journal={Journal of the European Economic Association},
  volume={8},
  number={1},
  pages={90--119},
  year={2010},
  publisher={Oxford University Press}
}

@article{hopkins2004running,
  title={Running to keep in the same place: Consumer choice as a game of status},
  author={Hopkins, Ed and Kornienko, Tatiana},
  journal={American Economic Review},
  volume={94},
  number={4},
  pages={1085--1107},
  year={2004}
}

@article{hopkins2023cardinal,
    author = {Hopkins, Ed},
    title = {Cardinal Sins? Conspicuous Consumption, Cardinal Status and Inequality},
    journal = {Journal of the European Economic Association},
    volume = {22},
    number = {5},
    pages = {2374-2413},
    year = {2024},
    month = {03},
    issn = {1542-4766},
    doi = {10.1093/jeea/jvae025},
    url = {https://doi.org/10.1093/jeea/jvae025},
    eprint = {https://academic.oup.com/jeea/article-pdf/22/5/2374/59646734/jvae025.pdf},
}

@incollection{jackson2015games,
  title={Games on networks},
  author={Jackson, Matthew O and Zenou, Yves},
  booktitle={Handbook of Game Theory with Economic Applications},
  volume={4},
  pages={95--163},
  year={2015},
  publisher={Elsevier}
}

@article{kahneman1979prospect,
  title={Prospect theory: an analysis of decision under risk},
  author={Kahneman, Daniel and Tversky, Amos},
  journal={Econometrica},
  volume={47},
  number={2},
  pages={263--292},
  year={1979},
  publisher={Oxford University Press}
}

@article{langtry2022keeping,
  title={Keeping up with "The Joneses": reference dependent choice with social comparisons},
  author={Langtry, Alastair},
  journal={American Economic Journal: Microeconomics},
  volume={15},
  number={3},
  pages={474--500},
  year={2022}
}

@article{ljungqvist2000tax,
  title={Tax policy and aggregate demand management under catching up with the Joneses},
  author={Ljungqvist, Lars and Uhlig, Harald},
  journal={American Economic Review},
  volume={90},
  number={3},
  pages={356--366},
  year={2000}
}

@article{loewenstein1989social,
  title={Social Utility and Decision Making in Interpersonal Contexts},
  author={Loewenstein, George F. and Thompson, Leigh and Bazerman, Max H.},
  journal={Journal of Personality and Social Psychology},
  volume={LVII},
  pages={426--441},
  year={1989}
}

@article{lopez-pintado2021far,
  title={Far above others},
  author={L{\'o}pez-Pintado, Dunia and Mel{\'e}ndez-Jim{\'e}nez, Miguel A.},
  journal={Journal of Economic Theory},
  volume={198},
  pages={105376},
  year={2021},
  doi={10.1016/j.jet.2021.105376}
}

@article{petrishcheva2023loss,
  title={Loss aversion in social image concerns},
  author={Petrishcheva, Vasilisa and Riener, Gerhard and Schildberg-H{\"o}risch, Hannah},
  journal={Experimental Economics},
  volume={26},
  number={3},
  pages={622--645},
  year={2023},
  publisher={Cambridge University Press}
}

@unpublished{sadler2023games,
  author={Sadler, Evan and Golub, Benjamin},
  title={Games on Endogenous Networks},
  year={2023}
}

@article{thaler1980toward,
  title={Toward a positive theory of consumer choice},
  author={Thaler, Richard},
  journal={Journal of Economic Behavior \& Organization},
  volume={1},
  number={1},
  pages={39--60},
  year={1980},
  publisher={Elsevier}
}

@article{tversky1992advances,
  title={Advances in prospect theory: Cumulative representation of uncertainty},
  author={Tversky, Amos and Kahneman, Daniel},
  journal={Journal of Risk and Uncertainty},
  volume={5},
  number={4},
  pages={297--323},
  year={1992},
  publisher={Springer}
}

@article{ushchev2020social,
  title={Social norms in networks},
  author={Ushchev, Philip and Zenou, Yves},
  journal={Journal of Economic Theory},
  volume={185},
  pages={104969},
  year={2020},
  publisher={Elsevier}
}

@article{ferrer2005income,
  title={Income and Well-being: An Empirical Analysis of the Comparison Income Effect},
  author={Ferrer-i-Carbonell, Ada},
  journal={Journal of Public Economics},
  volume={89},
  number={5-6},
  pages={997--1019},
  year={2005}
}

@article{brown-metaanalysis,
Author = {Brown, Alexander L. and Imai, Taisuke and Vieider, Ferdinand M. and Camerer, Colin F.},
Title = {Meta-analysis of Empirical Estimates of Loss Aversion},
Journal = {Journal of Economic Literature},
Volume = {62},
Number = {2},
Year = {2024},
Month = {June},
Pages = {485–516},
DOI = {10.1257/jel.20221698},
URL = {https://www.aeaweb.org/articles?id=10.1257/jel.20221698}}

@article{vendrik2007happiness,
  title={Happiness and loss aversion: Is utility concave or convex in relative income?},
  author={Vendrik, Maarten CM and Woltjer, Geert B},
  journal={Journal of Public Economics},
  volume={91},
  number={7-8},
  pages={1423--1448},
  year={2007},
  publisher={Elsevier}
}

@article{leites2022effect,
  title={The effect of relative income concerns on life satisfaction: Relative deprivation and loss aversion},
  author={Leites, Mart{\'\i}n and Ramos, Xavier},
  journal={Journal of Happiness Studies},
  volume={23},
  number={7},
  pages={3485--3515},
  year={2022},
  publisher={Springer}
}

@article{clark1996satisfaction,
  title={Satisfaction and comparison income},
  author={Clark, Andrew E and Oswald, Andrew J},
  journal={Journal of Public Economics},
  volume={61},
  number={3},
  pages={359--381},
  year={1996},
  publisher={Elsevier}
}

@article{senik2009direct,
  title={Direct evidence on income comparisons and their welfare effects},
  author={Senik, Claudia},
  journal={Journal of Economic Behavior \& Organization},
  volume={72},
  number={1},
  pages={408--424},
  year={2009},
  publisher={Elsevier}
}

@article{gill2012structural,
  title={A structural analysis of disappointment aversion in a real effort competition},
  author={Gill, David and Prowse, Victoria},
  journal={American Economic Review},
  volume={102},
  number={1},
  pages={469--503},
  year={2012},
  publisher={American Economic Association}
}

@techreport{bramoulle2024status,
  title = {Status Consumption in Networks: A Reference Dependent Approach},
  author = {Bramoull{\'e}, Yann and Ghiglino, Christian},
  year = {2024},
  institution = {Cambridge Working Papers in Economics},
  number = {2414},
  doi = {10.17863/CAM.107461}
}

@article{FehrCharness2025,
  author  = {Fehr, Ernst and Charness, Gary},
  title   = {Social Preferences: Fundamental Characteristics and Economic Consequences},
  journal = {Journal of Economic Literature},
  year    = {2025},
  volume  = {63},
  number  = {2},
  pages   = {440--514},
  doi     = {10.1257/jel.20241391}
}

@article{FehrSchmidt1999,
  author  = {Fehr, Ernst and Schmidt, Klaus M.},
  title   = {A Theory of Fairness, Competition, and Cooperation},
  journal = {The Quarterly Journal of Economics},
  year    = {1999},
  volume  = {114},
  number  = {3},
  pages   = {817--868},
  doi     = {10.1162/003355399556151}
}

@article{bursztyn2018status,
  title   = {Status Goods: Experimental Evidence from Platinum Credit Cards},
  author  = {Bursztyn, Leonardo and Ferman, Bruno and Fiorin, Stefano and Kanz, Martin and Rao, Gautam},
  journal = {Quarterly Journal of Economics},
  volume  = {133},
  number  = {3},
  pages   = {1561--1595},
  year    = {2018},
  publisher = {Oxford University Press},
  doi     = {10.1093/qje/qjy013}
}

@article{ClingingsmithSheremeta2018,
  author  = {Clingingsmith, David and Sheremeta, Roman M.},
  title   = {Status and the Demand for Visible Goods: Experimental Evidence on Conspicuous Consumption},
  journal = {Experimental Economics},
  year    = {2018},
  volume  = {21},
  number  = {4},
  pages   = {877--904},
  doi     = {10.1007/s10683-017-9529-0}
}

@article{BanuriNguyen2023,
  author  = {Banuri, Sheheryar and Nguyen, Trong-Anh},
  title   = {Status, Competition, and Performance: Experimental Evidence},
  journal = {Journal of Economic Behavior \& Organization},
  year    = {2023},
  volume  = {205},
  pages   = {237--255},
  doi     = {10.1016/j.jebo.2022.11.012}
}

@article{Bernheim1994,
  author  = {Bernheim, B. Douglas},
  title   = {A Theory of Conformity},
  journal = {Journal of Political Economy},
  year    = {1994},
  volume  = {102},
  number  = {5},
  pages   = {841--877},
  doi     = {10.1086/261957}
}

@article{AndreoniBernheim2009,
  author  = {Andreoni, James and Bernheim, B. Douglas},
  title   = {Social Image and the 50–50 Norm: A Theoretical and Experimental Analysis of Audience Effects},
  journal = {Econometrica},
  year    = {2009},
  volume  = {77},
  number  = {5},
  pages   = {1607--1636},
  doi     = {10.3982/ECTA7384}
}

@article{carroll1997comparison,
  title={Comparison utility in a growth model},
  author={Carroll, Christopher D and Overland, Jody and Weil, David N},
  journal={Journal of economic growth},
  volume={2},
  number={4},
  pages={339--367},
  year={1997},
  publisher={Springer}
}

@article{akerlof1997social,
  title={Social distance and social decisions},
  author={Akerlof, George A},
  journal={Econometrica: Journal of the Econometric Society},
  pages={1005--1027},
  year={1997},
  publisher={JSTOR}
}

@incollection{bisin2011economics,
  title={The economics of cultural transmission and socialization},
  author={Bisin, Alberto and Verdier, Thierry},
  booktitle={Handbook of social economics},
  volume={1},
  pages={339--416},
  year={2011},
  publisher={Elsevier}
}

@article{robson1992status,
  title={Status, the distribution of wealth, private and social attitudes to risk},
  author={Robson, Arthur J},
  journal={Econometrica: Journal of the Econometric Society},
  pages={837--857},
  year={1992},
  publisher={JSTOR}
}

@article{moldovanu2007contests,
  title={Contests for status},
  author={Moldovanu, Benny and Sela, Aner and Shi, Xianwen},
  journal={Journal of political Economy},
  volume={115},
  number={2},
  pages={338--363},
  year={2007},
  publisher={The University of Chicago Press}
}

@article{stark1984rural,
  title={Rural-to-urban migration in LDCs: a relative deprivation approach},
  author={Stark, Oded},
  journal={Economic Development and Cultural Change},
  volume={32},
  number={3},
  pages={475--486},
  year={1984},
  publisher={University of Chicago Press}
}

@article{bolton2000erc,
  title={ERC: A theory of equity, reciprocity, and competition},
  author={Bolton, Gary E and Ockenfels, Axel},
  journal={American economic review},
  volume={91},
  number={1},
  pages={166--193},
  year={2000},
  publisher={American Economic Association}
}

@article{zenoupeer,
author = {Boucher, Vincent and Rendall, Michelle and Ushchev, Philip and Zenou, Yves},
title = {Toward a General Theory of Peer Effects},
journal = {Econometrica},
volume = {92},
number = {2},
pages = {543-565},
  year={2024},
doi = {https://doi.org/10.3982/ECTA21048},
url = {https://onlinelibrary.wiley.com/doi/abs/10.3982/ECTA21048},
eprint = {https://onlinelibrary.wiley.com/doi/pdf/10.3982/ECTA21048},
}

@incollection{ledyard1995public,
  title = {Public Goods: A Survey of Experimental Research},
  author = {Ledyard, John O.},
  booktitle = {The Handbook of Experimental Economics},
  editor = {Kagel, John H. and Roth, Alvin E.},
  pages = {111--194},
  year = {1995},
  publisher = {Princeton University Press},
  address = {Princeton, NJ}
}
